\documentclass[prodmode,acmtoms]{acmsmall} % Aptara syntax

\usepackage{psfrag}
\usepackage{amsbsy}
\usepackage{amsmath}
\usepackage{amssymb}
\usepackage[ruled]{algorithm2e}

\newcommand{\mylab}[3]{\raisebox{#2}[0mm][0mm]{%
\makebox[0mm][l]{\hspace*{#1}\textbf{#3}}}}

\def\drawline#1#2{\raise 2.5pt\vbox{\hrule width #1pt height #2pt}}

\def\circle{$\circ$\nobreak }

\def\trian{\raise 1.25pt\hbox{$\scriptstyle\triangle$}\nobreak}

\def\dtrian{\raise 1.25pt\hbox%
{$\scriptscriptstyle\bigtriangledown$}\nobreak}

\def\squar{\raise 1.25pt\hbox{$\scriptstyle\Box$}\nobreak}

\def\diamon{\raise 1.25pt\hbox{$\scriptstyle\diamond$}\nobreak}

\def\beq{\begin{equation}}
\def\eeq{\end{equation}}

\def\aaa{{\it a}}
\def\bbb{{\it b}}
\def\ccc{{\it c}}
\def\ddd{{\it d}}

\def\citalajim03{Del \'Alamo \& Jim\'enez (2003)}
\renewcommand{\algorithmcfname}{ALGORITHM}
\SetAlFnt{\small}
\SetAlCapFnt{\small}
\SetAlCapNameFnt{\small}
\SetAlCapHSkip{0pt}
\IncMargin{-\parindent}

\acmVolume{X}
\acmNumber{X}
\acmArticle{XX}
\acmYear{2015}
\acmMonth{4}

\begin{document}

% Page heads
\markboth{A. Lozano-Dur\'an et al.}{An efficient algorithm to compute
  the genus of discrete surfaces and applications to turbulent
  flows}

% Title portion
\title{An efficient algorithm to compute the genus of discrete surfaces and applications to turbulent flows}
\author{ADRI\'AN LOZANO-DUR\'AN 
\affil{Universidad Polit\'ecnica de Madrid}
GUILLEM BORRELL 
\affil{Universidad Polit\'ecnica de Madrid}}

\begin{abstract}
  A simple and efficient algorithm to numerically compute the genus of
  surfaces of three-dimensional objects using the Euler characteristic
  formula is presented. The algorithm applies to objects obtained by
  thresholding a scalar field in a structured-collocated grid, and
  does not require any triangulation of the data. This makes the
  algorithm fast, memory-efficient and suitable for large
  datasets. Applications to the characterization of complex surfaces
  in turbulent flows are presented to illustrate the method.
\end{abstract}

\category{G.4}{Mathematical Software}{Algorithm design and analysis}
\category{I.3}{Computational Geometry and Object Modeling}{}
\category{J.2}{Physical Science and Engineering}{Physics}

\terms{Algorithms}

\keywords{genus, Euler characteristic, voxels, turbulence, coherent
  structures, turbulent/non-turbulent interface}

\acmformat{Adri\'an Lozano-Dur\'an and Guillem Borrell 2015.  An
  efficient algorithm to compute the genus of discrete surfaces and
  applications to turbulent flows}

\begin{bottomstuff}
This work was supported by the Computational Fluid Mech. Lab. headed
by Javier Jim\'enez, the European Research Council, under grant
ERC-2010.AdG-20100224. A.  Lozano--Dur\'an was supported by an FPI
fellowship from the Spanish Ministry of Education and Science, and
ERC.
\end{bottomstuff}

\maketitle

%==================================================================%
\section{Introduction}\label{sec:intro}
%==================================================================%

We present a fast and memory-efficient algorithm to numerically
compute the topological genus of all the surfaces associated with
three-dimensional objects in a discrete space. The paper is aimed at
the turbulence community interested in the topology of
three-dimensional entities in turbulent flows such as coherent
structures \cite{ala:jim:zan:mos:2006,loz:flo:jim:2012} or
turbulent/non-turbulent interfaces \cite{sil:2014}.
\citeN{Konkle:2003} describes fast methods for computing the genus of
triangulated surfaces which is usually a time and memory-consuming
process. Our algorithm does not rely on triangulation
\cite{tor:yon:2002,che:ron:2010,aya:ver:cru:2012,cru:aya:2013} and is
adapted to exploit the structured-collocated grid commonly used in the
largest direct numerical simulations of turbulent flows
\cite{kaneda:2003,hoyas:2008,sillero:2013}.  Our goal is to provide a
clear and easy description of the algorithm and sample codes. More
examples in Fortran and Python are available at \citeN{torroja}.

% definition of genus
The genus is a topologically invariant property of a surface defined
as the largest number of non-intersecting simple closed curves that
can be drawn on the surface without separating it.  The genus is
negative when applied to a group of several isolated surfaces, since
it is considered that no closed curves are required to separate
them. Both spheres and discs have genus zero, while a torus has genus
one. On the other hand, two separated spheres or the surfaces defined
by a sphere shell (or sphere with an internal cavity) has genus minus
one.  For a set of objects in a given region, the genus is equal to
the \emph{number of holes} - \emph{number of objects} - \emph{number
  of internal cavities}+1. The concept is also defined for higher
dimensions but the present work is restricted to two-dimensional
surfaces embedded in a three-dimensional space. In Integral Geometry,
the genus is part of a larger set of Galilean invariants called
Minkowski functionals which characterize the global aspects of a
structure in a $n$-dimensional space.  The genus is also closely
related to the Betti numbers, and more details can be found in
\citeN{tho:96}.

% genus in cosmology
Regarding its applications, the genus has proven to be very useful to
characterize a wide variety of structures in many fields, for
instance, in cosmology and related cosmic microwave background studies
\cite{ein:2007}.  The large-scale structure of the universe has been
studied over the years through analyses of the distribution of
galaxies in three dimensions using the genus for characterizing its
topology
\cite{got:86,got:87,got:89,ham:86,vog:94,mec:buc:wag:94,par:2005a,par:2005b}.
For a given threshold of the galaxy density, an isosurface separating
higher and lower density regions is defined and the genus of such
contour evaluated. This allows to compare the topology observed with
that expected for Gaussian random phase initial conditions
\cite{gut:81,lin:83}. In all these applications, the computation of
the genus was performed by calculating the discrete integrated
Gaussian curvatures \cite{got:86,che:ron:2010} following the Fortran
algorithm by \citeN{wei:88} based on the Gauss-Bonnet theorem.  As we
will show in section \ref{sec:algorithm}, the present method does not
rely on computing any curvatures.

% genus in medical research
Other applications are oriented to medical and biological areas and
use the genus of surfaces or three-dimensional objects. For example,
to compute adenine properties in the biochemistry field
\cite{Konkle:2003} and to evaluate the osteoporosis degree of mice
femur \cite{bad:2003} or human vertebrae \cite{odg:1993}.

% genus in turbulence
The Minkowski functionals have recently been introduced in the study
of turbulent flows through the so called shapefinders
\cite{sah:sat:shan:98}. \citeN{leu:swa:dav:2012} studied the
topological properties of enstrophy isosurfaces in isotropic
turbulence by filtering the data at different scales and computing
structures of high enstrophy together with its corresponding Minkowski
functionals. The geometry of the educed objects was then classified
with two non-dimensional quantities, `planarity' and `filamentarity',
which measure the shape of the structures.

In a recent work, \citeN{bor:jim:2013} followed an strategy based on
the genus to decide optimal thresholds in turbulent/non-turbulent
interfaces extracted from numerical data.  Several surfaces were
obtained by thresholding the fluctuating enstrophy field in a
turbulent boundary layer and their associated genus was used as an
indicator of the complexity of the interface. This topological
description was crucial to decide the range of thresholds where a
vorticity isocontour can be considered a turbulent/non-turbulent
interface.

The rest of the paper is organized as follows.  Important definitions
are provided in section \ref{sec:definitions}.  The algorithm to
compute the genus is described in section \ref{sec:algorithm}.  An
alternative method is presented in \S\ref{sec:aalgorithm}, and
validated with the previous one in section \ref{sec:validation}, which
also contains some scalability tests.  Two applications to turbulent
flows are shown in \S\ref{sec:applications}.  Finally, conclusions are
offered in section \ref{sec:conclusions}.\\

%==================================================================%
\section{Definitions}\label{sec:definitions}
%==================================================================%

% initial field
We will first introduce the definitions of object, voxel, surface,
hole, cavity, and genus.  The starting point is a discrete
three-dimensional scalar field, $\phi=\phi(i,j,k)$, with $i=1,..,n_x$,
$j=1,..,n_y$ and $k=1,..,n_z$, where $n_x$, $n_y$ and $n_z$ are the
number of grid points in each direction respectively, separated by a
grid spacing $\delta$. Given a thresholding value $\alpha$, we define
the points belonging to the three-dimensional objects as those
satisfying
\begin{equation}\label{eq:thres}
\phi(i,j,k)>\alpha,
\end{equation}
which can be expressed a scalar field $A=A(i,j,k)$ whose values are
equal to 1 at $(i,j,k)$ if relation (\ref{eq:thres}) is satisfied, and
0 otherwise. The latter is refer to as an empty region.

\begin{figure}
\centerline{
\mylab{-0.8cm}{3.8cm}{(\aaa)}
\includegraphics[width=0.35\textwidth]{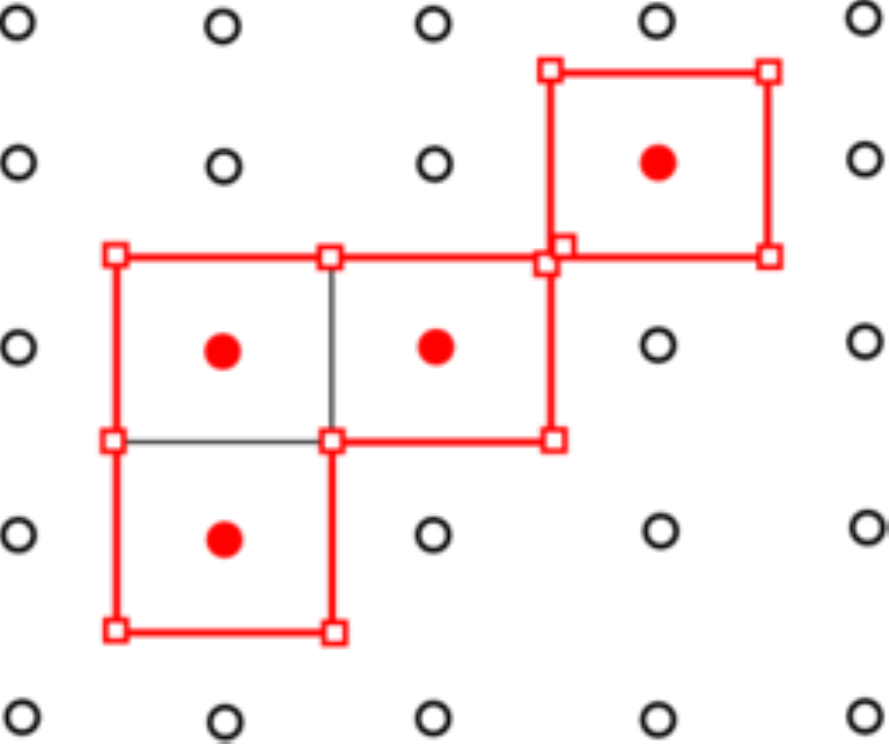}
\hspace{0.5cm}
\mylab{0.7cm}{3.8cm}{(\bbb)}
\includegraphics[width=0.45\textwidth]{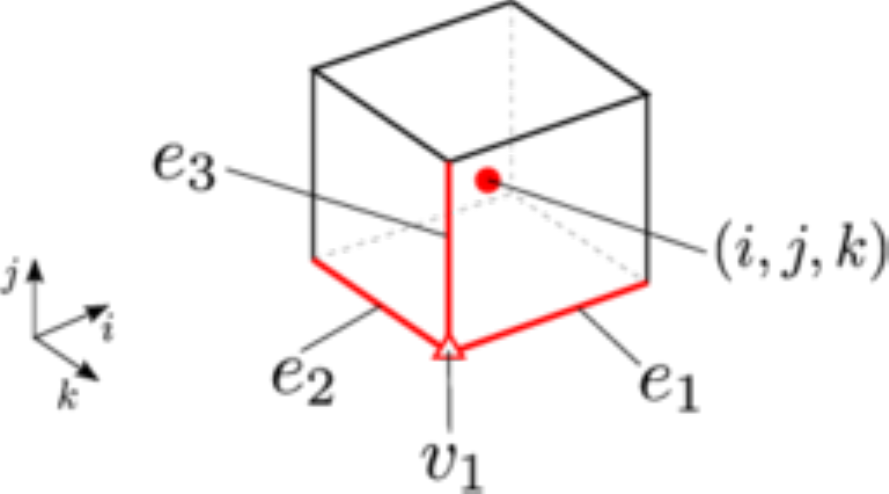}
}
\caption{ (a) Two-dimensional example of a structured-collocated grid
  defined by the open and closed circles. Points satisfying relation
  (\ref{eq:thres}) are red closed circles and their corresponding
  voxels are solid lines. The exterior edges are the thicker lines
  colored in red and the exterior vertices are marked by squares.
  Objects are built by connecting orthogonal neighboring points with
  $A=1$, which results in two objects in this particular example.  (b)
  Voxel around the grid point $(i,j,k)$. Only edges $e_1$, $e_2$ and
  $e_3$ (red lines), and vertex $v_1$ (triangle) are taken into
  account to compute the number of exterior vertices and edges
  corresponding to the voxel centered at $(i,j,k)$.
\label{fig:algorithm:vcube}}
\end{figure}

% objects
Three-dimensional individual objects in $A$ are constructed by
connecting neighboring points with value 1. Figure
\ref{fig:algorithm:vcube}(a) shows a two-dimensional example.
Connectivity is defined in terms of the six orthogonal neighbors in
the grid, usually called 6-connectivity. Points contiguous in oblique
directions are not directly connected, although they may become so
indirectly through connections with other points.  This remark is
important since the 6-connectivity is built-in in the algorithm and,
for instance, the number of objects in the example shown in Figure
\ref{fig:algorithm:vcube}(a) is not one but two.
%
%_________________________________________________________________%
%_________________________________________________________________%

% voxel, surface, holes and cavities
We define the voxel associated with $A(i,j,k)=1$ as the cube centered
at $(i,j,k)$ and with edge length equal to $\delta$ (see Figure
\ref{fig:algorithm:vcube}b). For a given object, its surface is
delimited by the exterior faces of its voxels, i.e., those facing
empty regions. In the 2D example shown in Figure
\ref{fig:algorithm:vcube}(a), the 1D `surface' is highlighted with red
lines. Actual 3D examples are shown in Figure
\ref{fig:validation:synthetic}. A hole is a empty region piercing the
object, as the torus in Figure \ref{fig:validation:synthetic}(a), and
a cavity an internal empty region which is locally not connected to
the exterior. The term handle will be used occasionally as a synonym
of hole, since they are topologically equivalent.

% genus and Euler characteristic
Our goal is to compute the genus of all the surfaces contained in $A$.
Mathematically, the genus $g$ is defined in terms of the Euler
characteristic $\chi$ via the relationship
\begin{equation}\label{eq:euler}
 \chi = 2 - 2g.
\end{equation}
The Euler characteristic can be calculated for continuous surfaces as
\begin{equation}\label{eq:euler_cont}
 \chi = \frac{1}{2\pi} \iint\limits_\Sigma K \mathrm{d}\Sigma,
\end{equation}
where $K$ is the Gaussian curvature of all the objects considered and
$\Sigma$ their area. However, we are more interested in the original
discrete form for polyhedral surfaces,
\begin{equation}\label{eq:euler_poly}
 \chi = F - E + V,
\end{equation}
where $F$, $E$, and $V$ are, respectively, the number of exterior
faces, edges and vertices of all the polyhedra. In this case, the
curvature can be considered to be located at the discrete edges, but
the calculations lead to the same results as
(\ref{eq:euler_cont}). The connection between the discrete and
continuous formulations is the Gauss-Bonnet theorem
\cite[p.\,243]{Chavel2006}. Intuitively, in terms of the elements
defined above, the genus is equal to the \emph{number of holes} -
\emph{number of objects} - \emph{number of internal cavities} $+
1$. \\

%==================================================================%
\section{Algorithm}\label{sec:algorithm}
%==================================================================%

% intro
The present algorithm exploits formula (\ref{eq:euler_poly}) and the
structured-collocated nature of the data to compute the genus of all
the surfaces contained in the three-dimensional space defined by the
scalar field $A$, without previous triangulation or calculation of the
Gaussian curvatures.  Note that this differs from other works which
compute the genus of the three-dimensional objects themselves
\cite{tor:yon:2002,che:ron:2010,aya:ver:cru:2012,cru:aya:2013}. The
method is conceived for large datasets of the order of $10^2$ GiB and
takes $A$ as input.

% idea and howto go through the domain
First, we provide a general description of the algorithm.  The key
idea is to place a voxel around every $(i,j,k)$ point with
$A(i,j,k)=1$, as the example shown in Figure
\ref{fig:algorithm:vcube}(b), and to create a virtual mesh using the
exterior elements of the resulting polyhedra. The term virtual is used
here in the sense that no actual faces, vertices or edges have to be
stored for each object, i.e., there is no actual structure in the code
to do so, in contrast to the standard meshes obtained by
triangulation, where these are saved in a file or in memory for all
the objects.  Figure \ref{fig:algorithm:vcube}(a) shows an example of
a virtual mesh in a two-dimensional case.

The way to proceed is then to compute the Euler characteristic of the
virtual mesh and thereafter the genus.  The value of $\chi$ is easily
calculated once the total number of exterior faces, vertices and edges
are known for all the objects within $A$.  To achieve so, three
variables $F$, $V$ and $E$ are used to store the total number of
exterior faces, vertices and edges, respectively, which are counted
looping once through the array $A$. At each $(i,j,k)$, a voxel is
placed if $A(i,j,k)=1$ and the counters $F$, $V$ and $E$ (initially
set to zero) are increased accordingly every time faces, edges and
vertices are identified as exterior. The selection of edges and
vertices taken into account at each $(i,j,k)$, shown in Figure
\ref{fig:algorithm:vcube}(b), is deliberately chosen to avoid counting
several times edges and vertices already considered.

To prevent any problems at the boundaries of the field $A$, the
original grid is extended by padding two extra planes of zeros at the
beginning and at the end of each dimension.  The new field will be
still called $A$, but now with dimensions $N_x=n_x+4$, $N_y=n_y+4$ and
$N_z=n_z+4$.  For simplicity, we consider that $A$ is fully loaded in
memory, but note that this is not required and it could be loaded in
small chunks or planes.

\begin{figure}
\centerline{
\mylab{0.9cm}{3.5cm}{(\aaa)}
\raisebox{-10pt}{\includegraphics[width=0.40\textwidth]{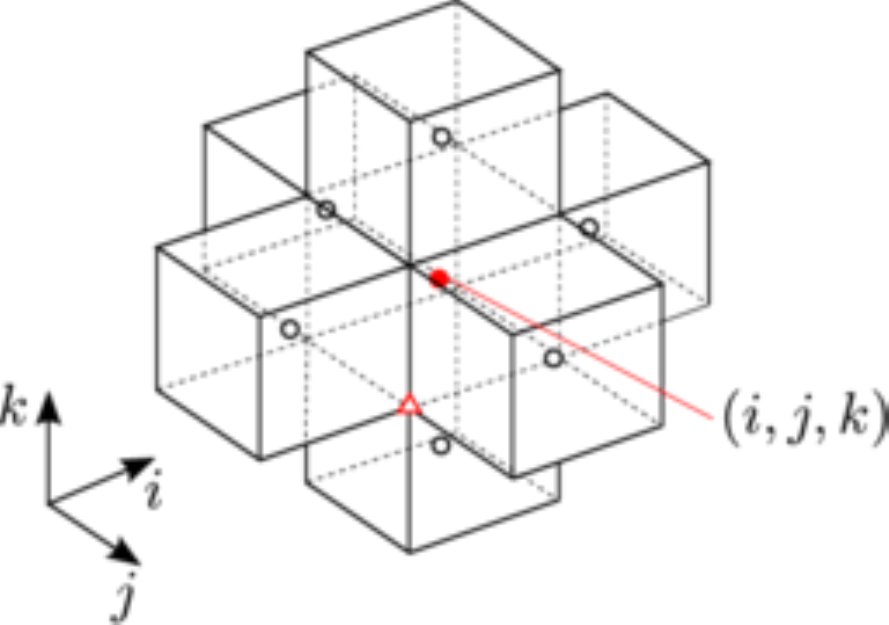}}
\hspace{0.2cm}
\mylab{0.1cm}{3.5cm}{(\bbb)}
\includegraphics[width=0.27\textwidth]{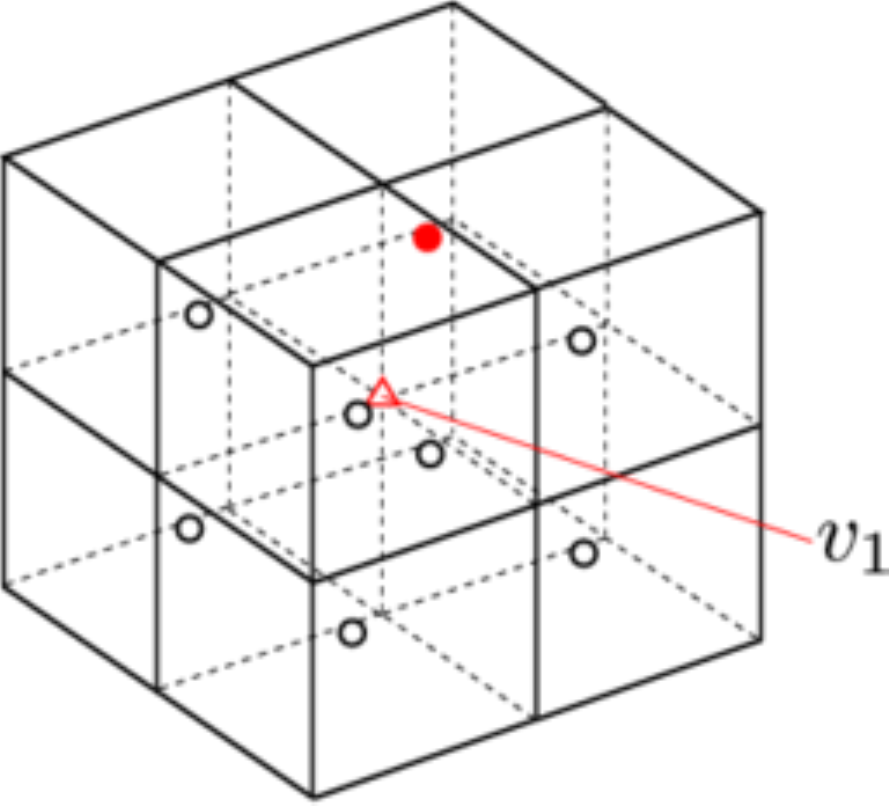}
\hspace{0.7cm}
\mylab{-0.1cm}{3.5cm}{(\ccc)}
\includegraphics[width=0.22\textwidth]{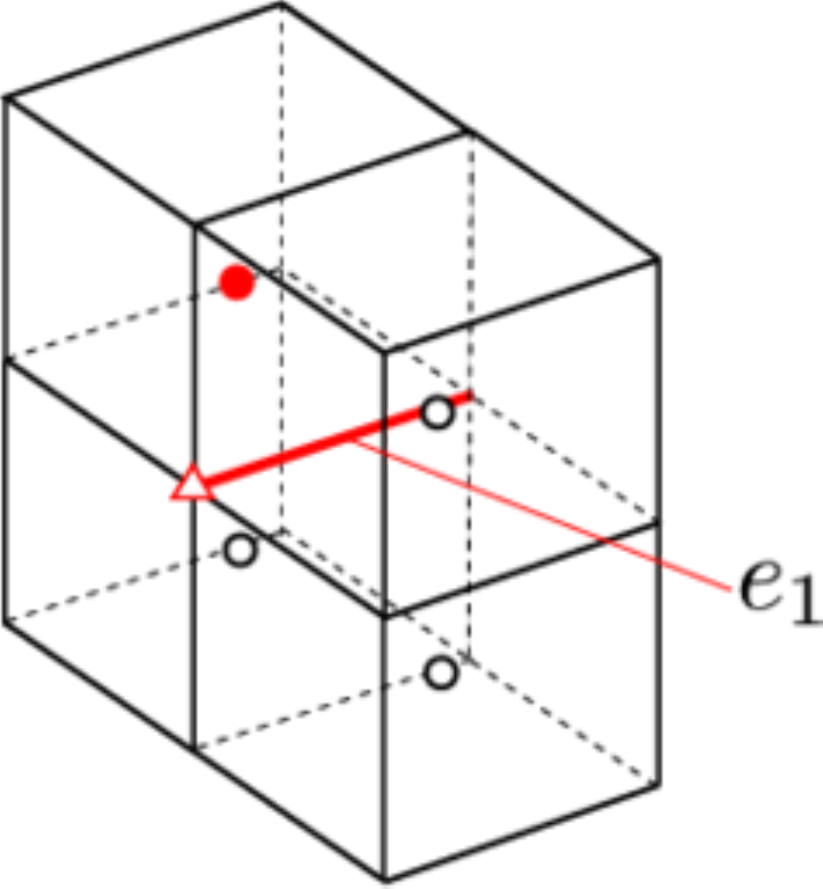}
}
\caption{Sketches of the neighboring voxels adjacent to voxel
  $(i,j,k)$ used to compute the number of exterior (a) faces, (b)
  vertices, and (c) edges.  In all plots, the closed red dot denotes
  the center of the voxel $(i,j,k)$ and the triangle the position of
  its vertex $v_1$ as defined in Figure
  \ref{fig:algorithm:vcube}(b). Case (c) is particularized to edge
  $e_1$ and similar configurations apply to edges $e_2$ and $e_3$ (see
  Figure \ref{fig:algorithm:vcube}b).
\label{fig:algorithm:cases}}
\end{figure}

% howto count faces, vertices edges
A more detailed description of the algorithm is now presented:
\begin{enumerate}
\item \emph{Initialize the variables}. $F, V$, and $E$ are integers
  containing the number of exterior faces, vertices and edges, and are
  initially set to zero.  For large cases, they must be double
  precision.  $A(i,j,k)$ is the array whose points are set to 1 if
  they belong to an object and to 0 otherwise.  $N_x,N_y$ and $N_z$
  are the sizes of $A$ after extending it.  $cube1$ and $cube2$ are
  auxiliary arrays of integers with dimensions $2\times2\times2$ and
  $2\times2$, respectively, and are used to store the slices of $A$
  shown in Figures \ref{fig:algorithm:cases}(b,c).

\item \emph{Loop through $i=2,..,N_x-1$, $j=2,..,N_y-1$,
  $k=2,..,N_z-1$}.  For each $(i,j,k)$ proceed as follows:
  \begin{enumerate}
  \item \emph{Count number of exterior faces}.  See algorithm
    \ref{alg:count_faces}. The six faces of the voxel at $(i,j,k)$ are
    considered, and its six neighbors are defined in Figure
    \ref{fig:algorithm:cases}(a).  For each neighboring voxel with
    coordinates $(i+\Delta i,j+\Delta j,k+\Delta k)$, $F$ is increased
    by one if $A(i,j,k)=1$ and $A(i+\Delta i,j+\Delta j,k+\Delta
    k)=0$. The possible values for $(i+\Delta i,j+\Delta j,k+\Delta
    k)$ are $(i-1,j,k)$, $(i+1,j,k)$, $(i,j-1,k)$, $(i,j+1,k)$,
    $(i,j,k-1)$ and $(i-1,j,k+1)$.
    %____________________________________________________________________%
    \begin{algorithm}[t]
    \SetAlgoNoLine 
    \KwIn{A,F}
    \KwOut{F}  
    \If{ A is equal to 1 at position (i,j,k)  } {
    \If{ A is equal to 0 at position (i-1,j,k)} { F $\leftarrow$ F + 1\;} 
    \If{ A is equal to 0 at position (i+1,j,k)} { F $\leftarrow$ F + 1\;}  
    \If{ A is equal to 0 at position (i,j-1,k)} { F $\leftarrow$ F + 1\;} 
    \If{ A is equal to 0 at position (i,j+1,k)} { F $\leftarrow$ F + 1\;} 
    \If{ A is equal to 0 at position (i,j,k-1)} { F $\leftarrow$ F + 1\;} 
    \If{ A is equal to 0 at position (i,j,k+1)} { F $\leftarrow$ F + 1\;} 
    }
    \caption{Count number of exterior faces at position $(i,j,k)$.}
    \label{alg:count_faces}
    \end{algorithm}
    %____________________________________________________________________%
% 
  \item \emph{Count number of exterior vertices}.  See algorithm
    \ref{alg:count_vertices}.  Only vertex $v_1$ in Figure
    \ref{fig:algorithm:vcube}(b) is considered, and its eight adjacent
    voxels are defined in Figure \ref{fig:algorithm:cases}(b).  $V$ is
    increased by one if any of the eight adjacent voxels has value 1,
    and any other value 0.  In some cases, $V$ must increase by a
    number $dV$ larger then 1 if some of the surrounding voxels are
    locally not connected.  The value of $dV$ is calculated by
    procedure $getconnected1$, which is discussed at the end of the
    section.
    %____________________________________________________________________%
    \begin{algorithm}[t]
    \SetAlgoNoLine 
    \KwIn{A,V}
    \KwOut{V}  
    cube1 $\leftarrow$ slice of A from $i-1$ to $i$, $j-1$ to $j$ and $k-1$ to $k$\;
    \If{any element in cube1 is 0 \and 
    any element in cube1 is 1}
    {
      V $\leftarrow$ V + getconnected1(cube1)\;
    }
    \caption{Count number of exterior vertices at position $(i,j,k)$.}
    \label{alg:count_vertices}
    \end{algorithm}
    %____________________________________________________________________%
%
  \item \emph{Count number of exterior edges}.  See algorithm
    \ref{alg:count_edges}. Only edges $e_1$, $e_2$ and $e_3$
    highlighted in Figure \ref{fig:algorithm:vcube}(b) are
    considered. The four adjacent voxels for edge $e_1$ are shown in
    Figure \ref{fig:algorithm:cases}(c).  $E$ is increased by one unit
    if any of the four adjacent voxels has value 1 and any other value
    0. In some cases, the edges have to be counted $dE$ times when the
    voxels around are not locally connected.  The increment $dE$ is
    computed by procedure $getconnected2$.
    %____________________________________________________________________%
    \begin{algorithm}[t]
    \SetAlgoNoLine 
    \KwIn{A,E}
    \KwOut{E}  
    cube2 $\leftarrow$ slice of A from $i-1$ to $i$, $j-1$ to $j$ and $k$\;
    \If{ any element in cube2 is 0 \and 
         any element in cube2 is 1}
         {
           E $\leftarrow$ E + getconnected2(cube2)\;
         }
    \caption{Count number of exterior edges at position $(i,j,k)$.}
    \label{alg:count_edges}
    \end{algorithm}
    %____________________________________________________________________%
    A similar algorithm applies to the other two edges $e_2$ and $e_3$
    shown in Figure \ref{fig:algorithm:vcube}(b).
\end{enumerate}
\item Finally the Euler characteristic and the genus are computed as
  $X = V - E + F$ and $G = (2 - X)/2$.
\end{enumerate}
%
%_________________________________________________________________%
%_________________________________________________________________%

% connectivity, getconnected
To complete the description of the algorithm, we now comment on the
procedures $getconnected1$ and $getconnected2$.  Some edges or
vertices has to be counted multiple times in order to be consistent
with the 6-connectivity of the voxels. An example is illustrated in
Figure \ref{fig:algorithm:vcube}(a). In contrast to other works
\cite{tor:yon:2002,aya:ver:cru:2012,cru:aya:2013}, this is achieved by
counting the number of local objects contained in the slices shown in
Figures \ref{fig:algorithm:cases}(b,c), i.e., the number of objects in
the $2\times2\times2$ sub-volume satisfying the 6-connectivity
disregarding any other connections outside the slide.  For example,
the sub-volume denoted as C41 in Figure \ref{fig:aalgorithm_sketches}
contains one local object, and C33 contains three.  Note that some
voxels may be locally disconnected but belong to the same object,
since they may connect indirectly through other voxels not considered
in the slide. The purpose of procedure $getconnected1$, defined in
algorithm \ref{alg:getconnected1}, is to compute the number of local
objects in the sub-volume shown in Figure
\ref{fig:algorithm:cases}(b), which can be easily obtained by any
labeling method like the Hoshen-Kopelman algorithm \cite{kop:76}. Note
that there is one degenerated case with a infinitesimally small hole,
shown in case C63 in Figure \ref{fig:aalgorithm_sketches}, where there
is only one object but the vertex must be considered twice.
%
%_____________________________________________________________________%
\begin{algorithm}[t]
\SetAlgoNoLine 
\KwIn{cube1} 
\KwOut{dV}  
\eIf{ cube1 is degenerated case C63 in Figure \ref{fig:aalgorithm_sketches}} 
    { dV $\leftarrow$ 2\;} 
    { 
      { dV $\leftarrow$ number of local objects in cube1\;} 
    }
\caption{Procedure getconnected1. It computes the increment $dV$ for V.}
\label{alg:getconnected1}
\end{algorithm}
%_____________________________________________________________________%
%

Procedure $getconnected2$ is presented in algorithm
\ref{alg:getconnected2} and follows the same idea.  In this case, the
only possible configuration to obtained more than one local object in
the slide shown in Figure \ref{fig:algorithm:cases}(c) is with two
voxels that do not share any face. In the rest of the cases, the
number of local objects is one.
%_____________________________________________________________________%
\begin{algorithm}[t]
\SetAlgoNoLine 
\KwIn{cube2} 
\KwOut{dE}
  \eIf {all the elements equal to 1 in cube2 are locally connected}{
     dE $\leftarrow$ 1\;
     }{
     dE $\leftarrow$ 2\;
     }
\caption{Procedure getconnected2. It computes the increment $dE$ for E.}
\label{alg:getconnected2}
\end{algorithm}
%_____________________________________________________________________%
\\

%==================================================================%
\section{Alternative algorithm}\label{sec:aalgorithm}
%==================================================================%

An alternative algorithm is introduced for the purpose of validating
the approach presented above. Conceptually, it follows the same ideas
discussed in section \ref{sec:algorithm} but relies on a pre-computed
table of cases as in the work by \citeN{tor:yon:2002}. The process
involves looping through all the vertices of the virtual grid,
counting vertices, faces and edges, but no effort is made to prevent
multiple counts of the last two, as opposed to the algorithm presented
in section \ref{sec:algorithm}. This results in an extra number of
faces and edges that is easily corrected by dividing the total number
of faces by 4 and of edges by 2, the reason being that each face and
edge contains 4 and 2 vertices, respectively.

\begin{figure}
\centerline{
  \includegraphics[width=0.1\textwidth]{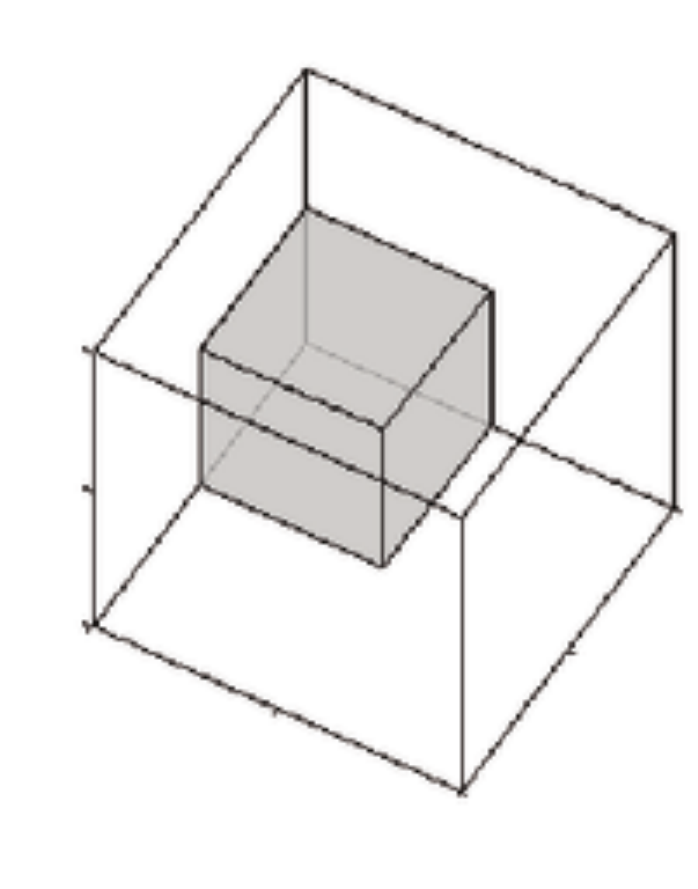}
  \mylab{-1.4cm}{-0.1cm}{$C_{11}$}
  \includegraphics[width=0.1\textwidth]{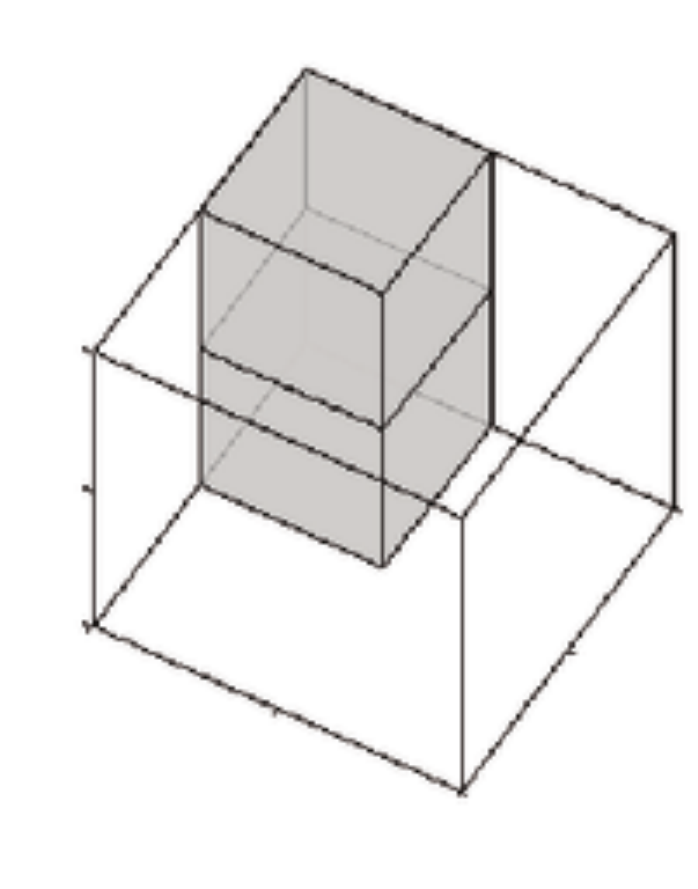}
  \mylab{-1.4cm}{-0.1cm}{$C_{21}$}
  \includegraphics[width=0.1\textwidth]{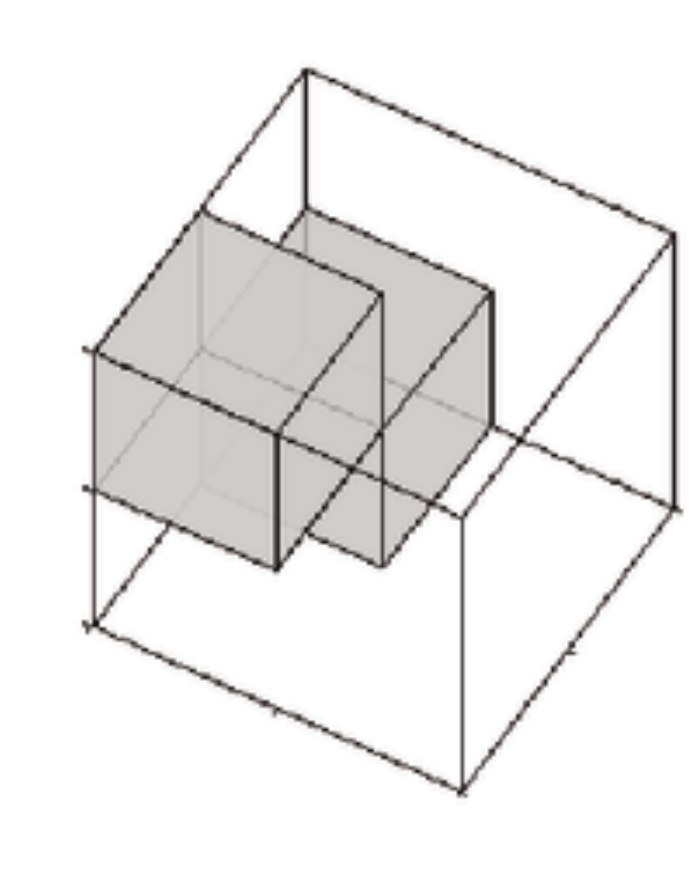}
  \mylab{-1.4cm}{-0.1cm}{$C_{22}$}
  \includegraphics[width=0.1\textwidth]{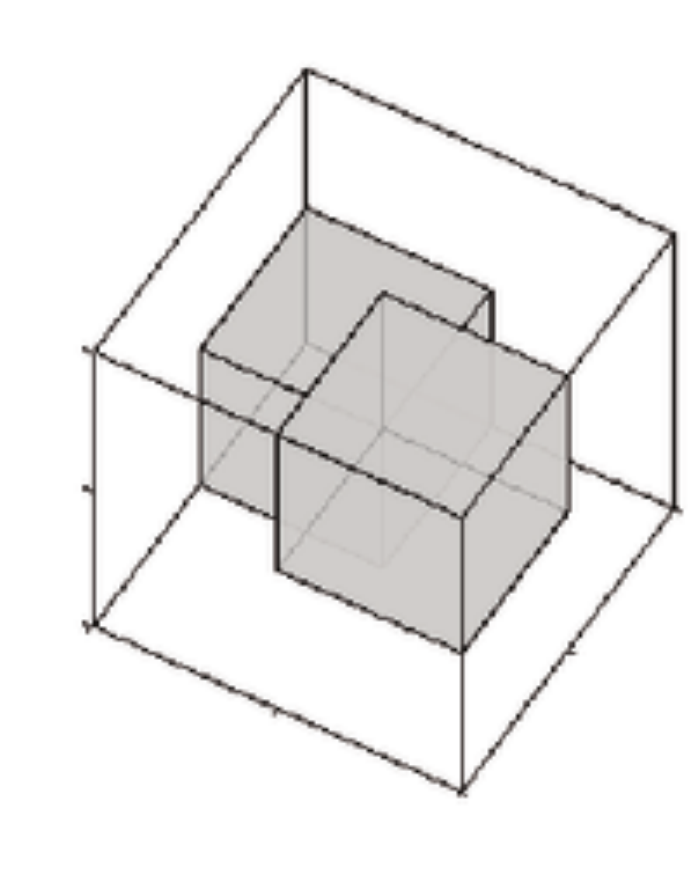}
  \mylab{-1.4cm}{-0.1cm}{$C_{23}$}
  \includegraphics[width=0.1\textwidth]{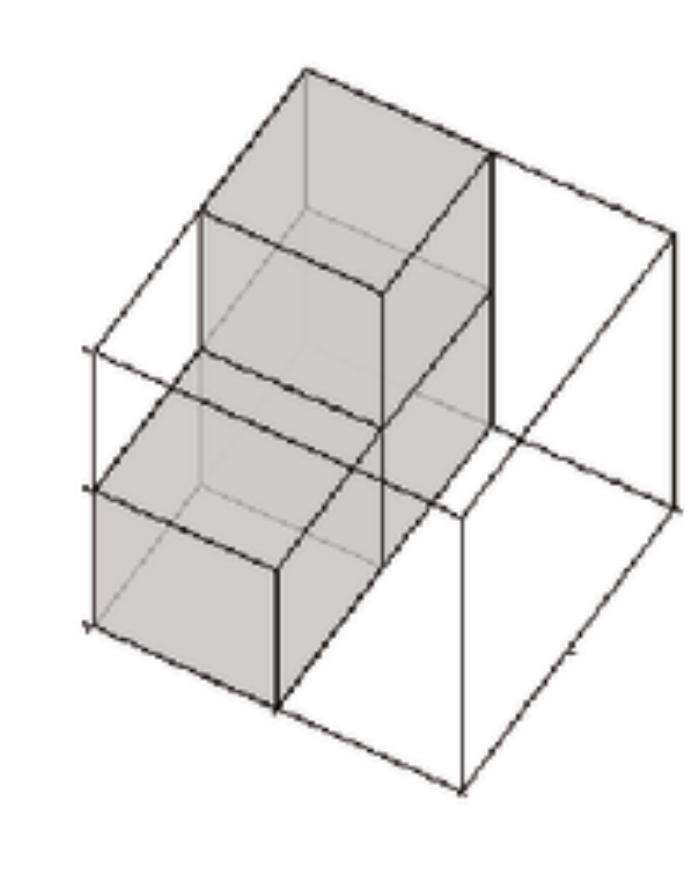}
  \mylab{-1.4cm}{-0.1cm}{$C_{31}$}
}
\vspace{0.5cm}
\centerline{
  \includegraphics[width=0.1\textwidth]{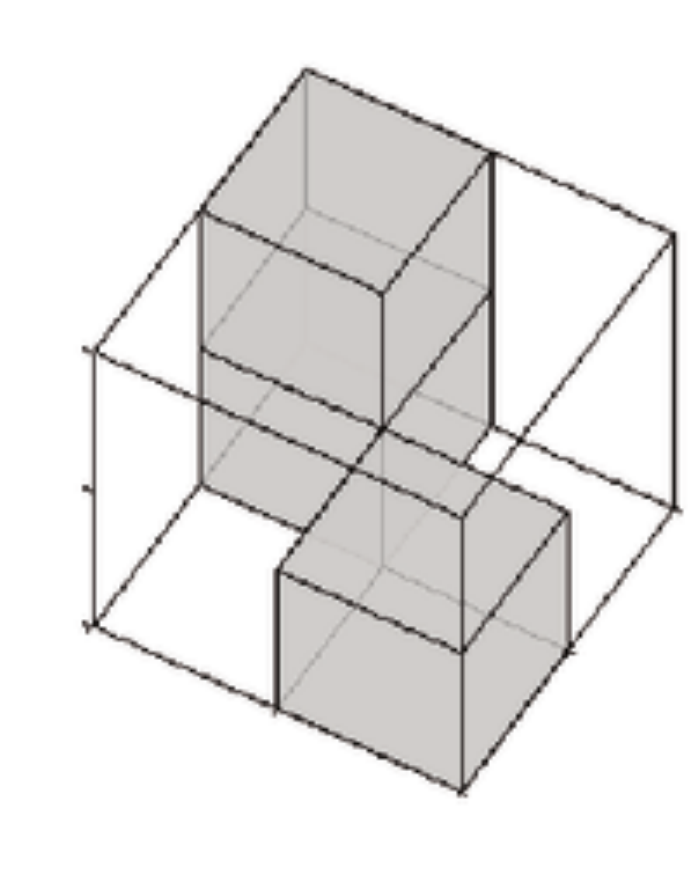}
  \mylab{-1.4cm}{-0.1cm}{$C_{32}$}
  \includegraphics[width=0.1\textwidth]{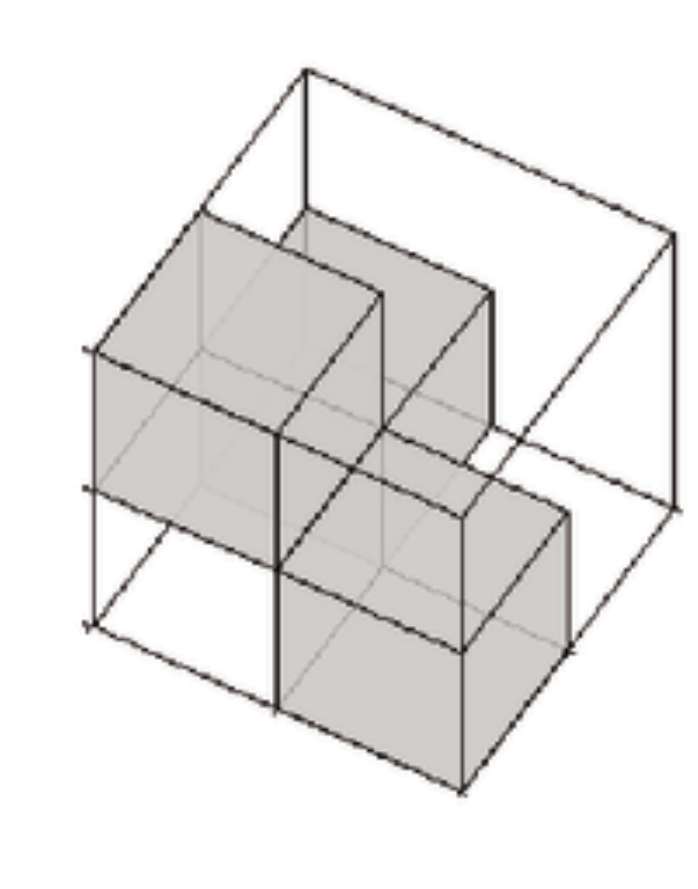}
  \mylab{-1.4cm}{-0.1cm}{$C_{33}$}
  \includegraphics[width=0.1\textwidth]{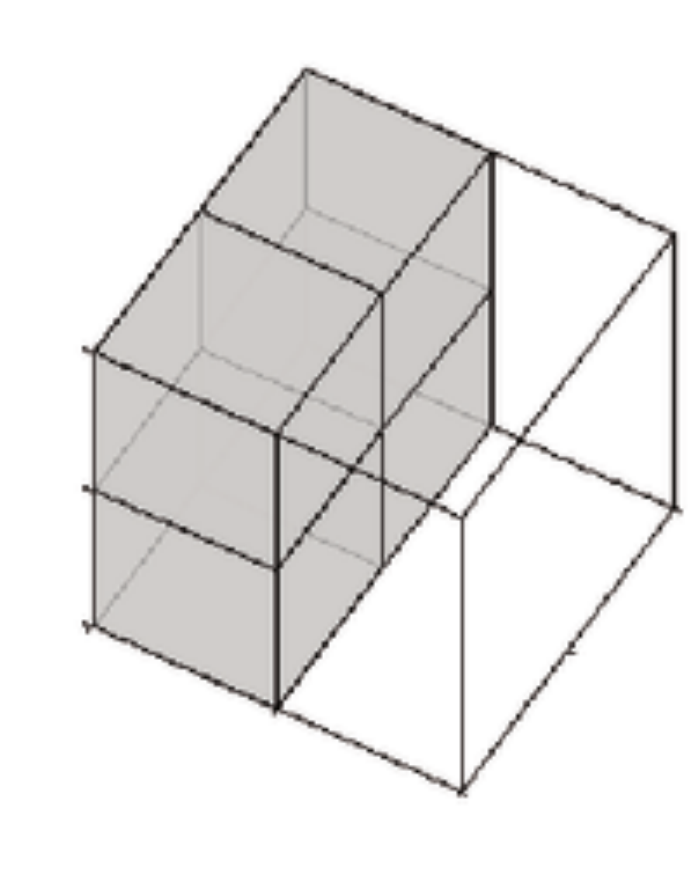}
  \mylab{-1.4cm}{-0.1cm}{$C_{41}$}
  \includegraphics[width=0.1\textwidth]{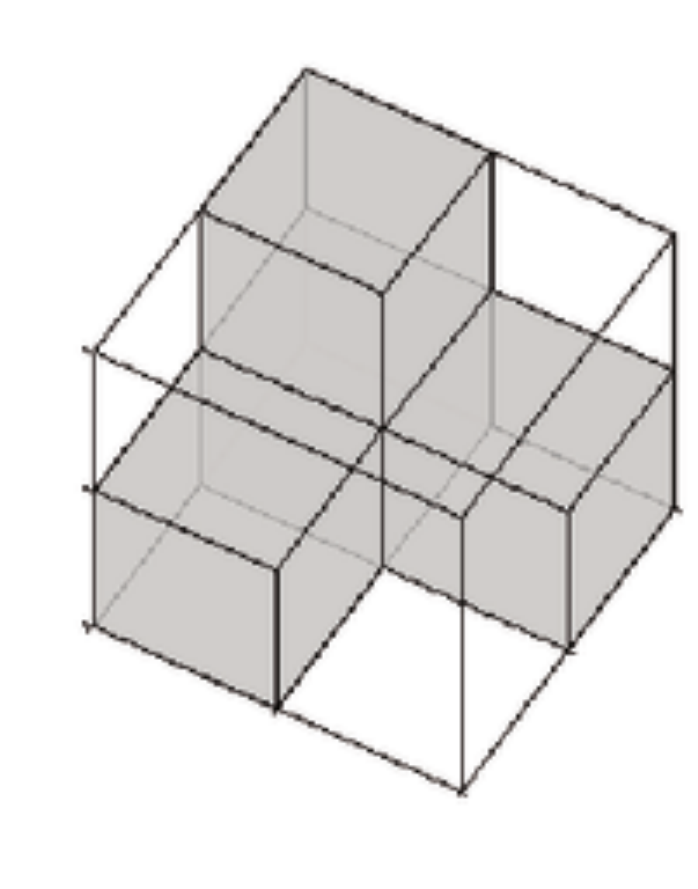}
  \mylab{-1.4cm}{-0.1cm}{$C_{42}$}
  \includegraphics[width=0.1\textwidth]{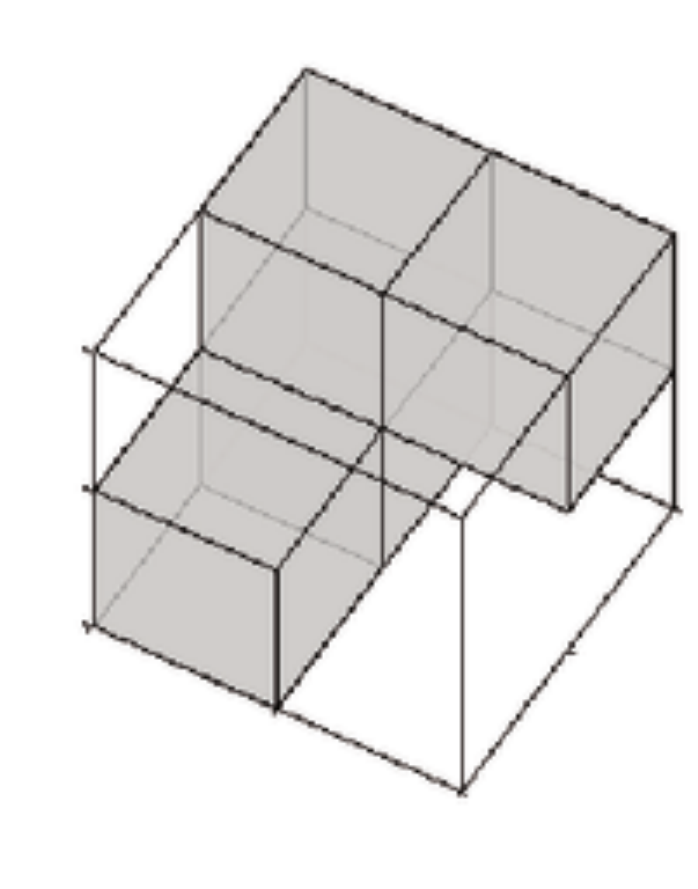}
  \mylab{-1.4cm}{-0.1cm}{$C_{43}$}
}
\vspace{0.5cm}
\centerline{
  \includegraphics[width=0.1\textwidth]{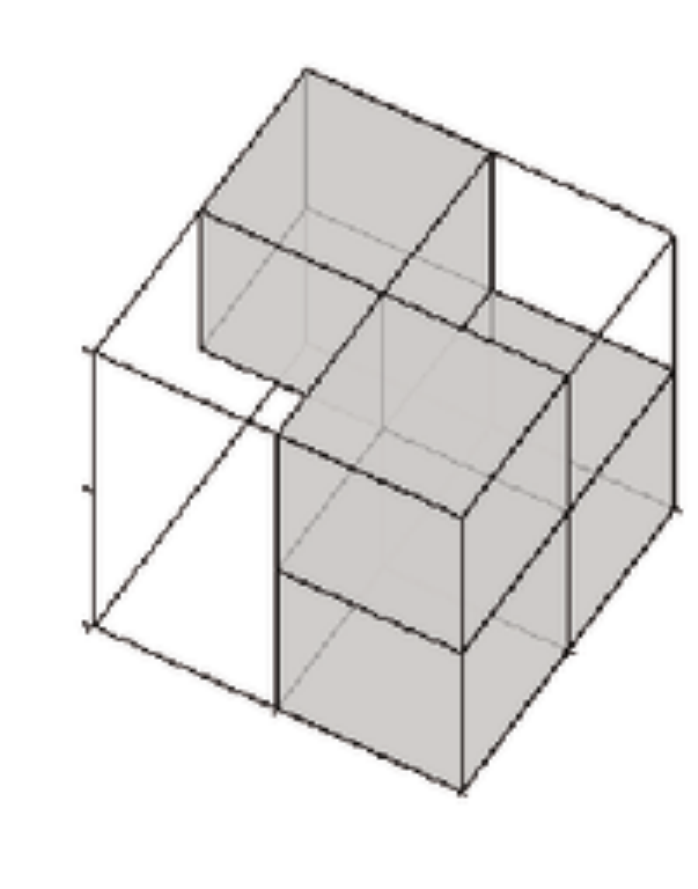}
  \mylab{-1.4cm}{-0.1cm}{$C_{44}$}
  \includegraphics[width=0.1\textwidth]{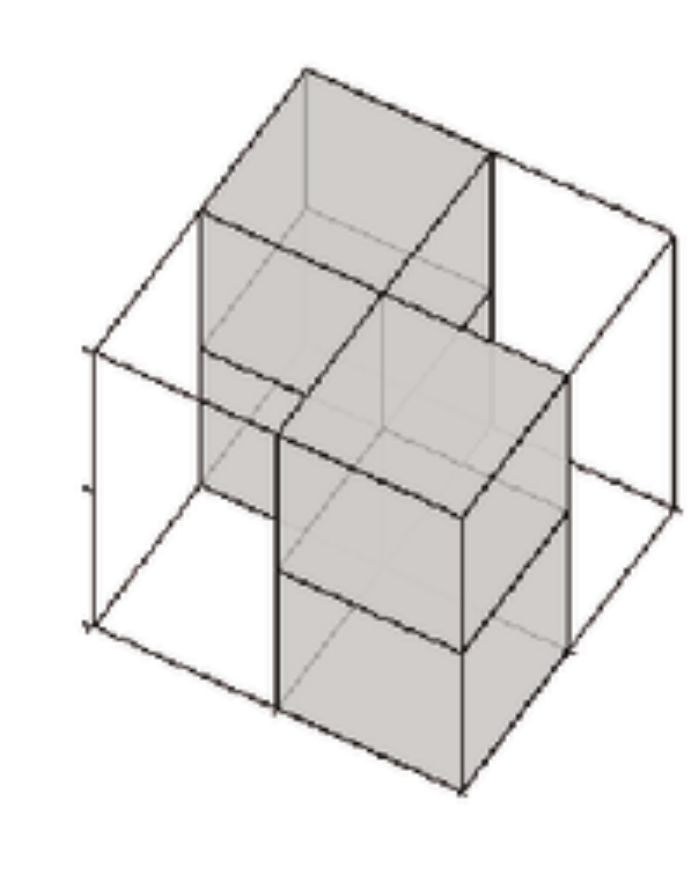}
  \mylab{-1.4cm}{-0.1cm}{$C_{45}$}
  \includegraphics[width=0.1\textwidth]{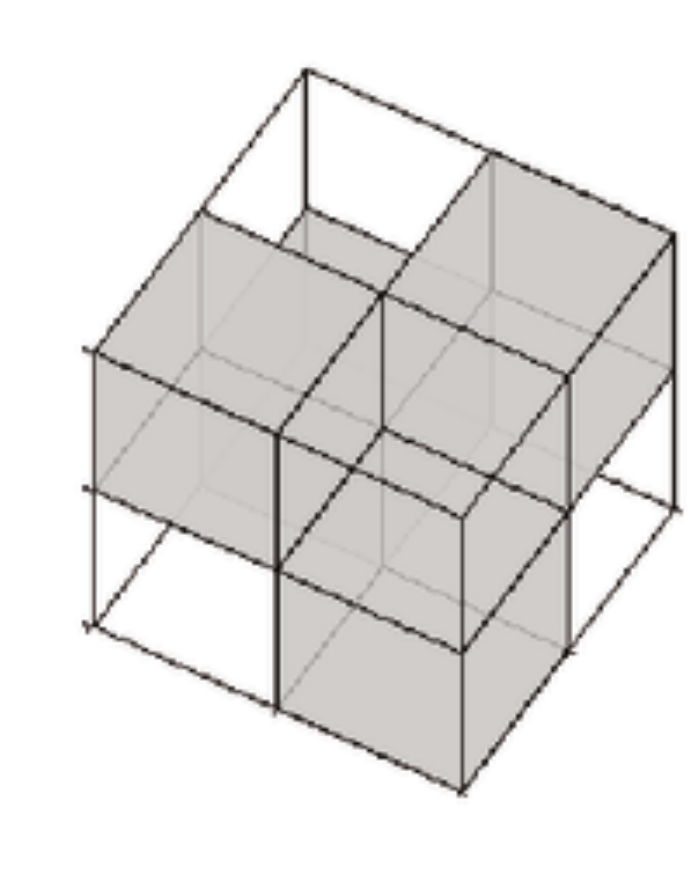}
  \mylab{-1.4cm}{-0.1cm}{$C_{46}$}
  \includegraphics[width=0.1\textwidth]{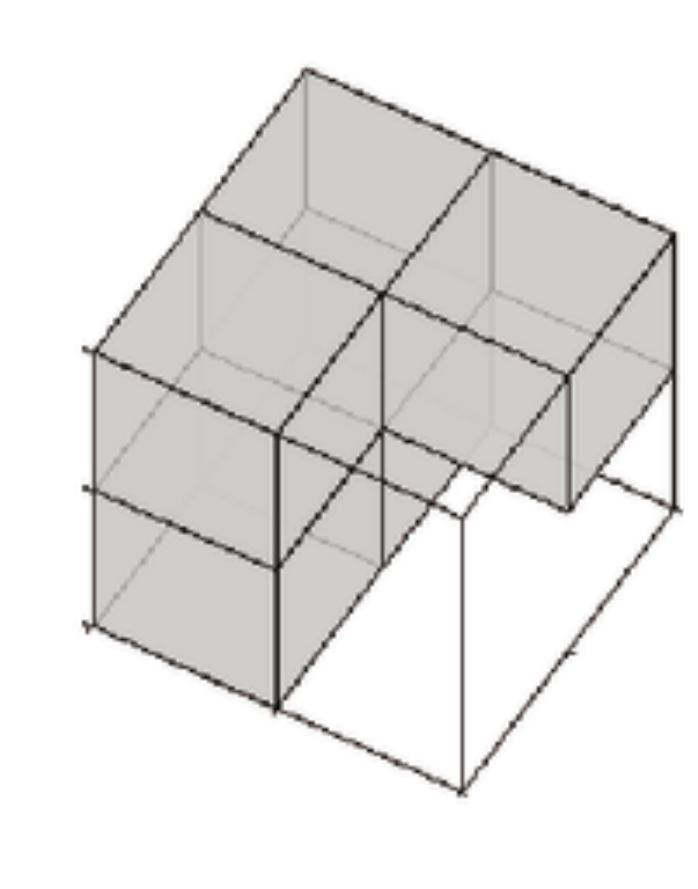}
  \mylab{-1.4cm}{-0.1cm}{$C_{51}$}
  \includegraphics[width=0.1\textwidth]{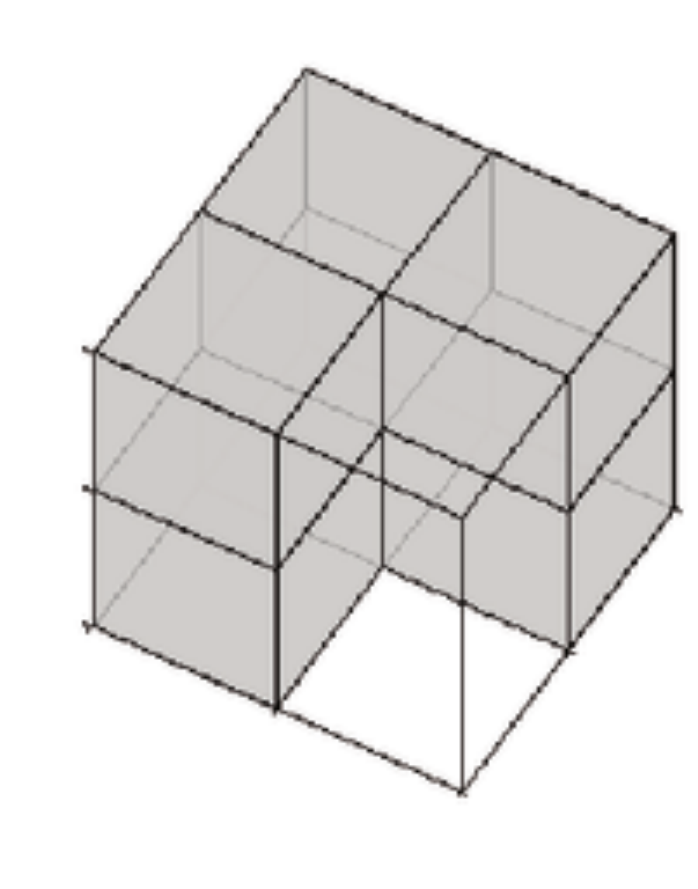}
  \mylab{-1.4cm}{-0.1cm}{$C_{52}$}
}
\vspace{0.5cm}
\centerline{
  \includegraphics[width=0.1\textwidth]{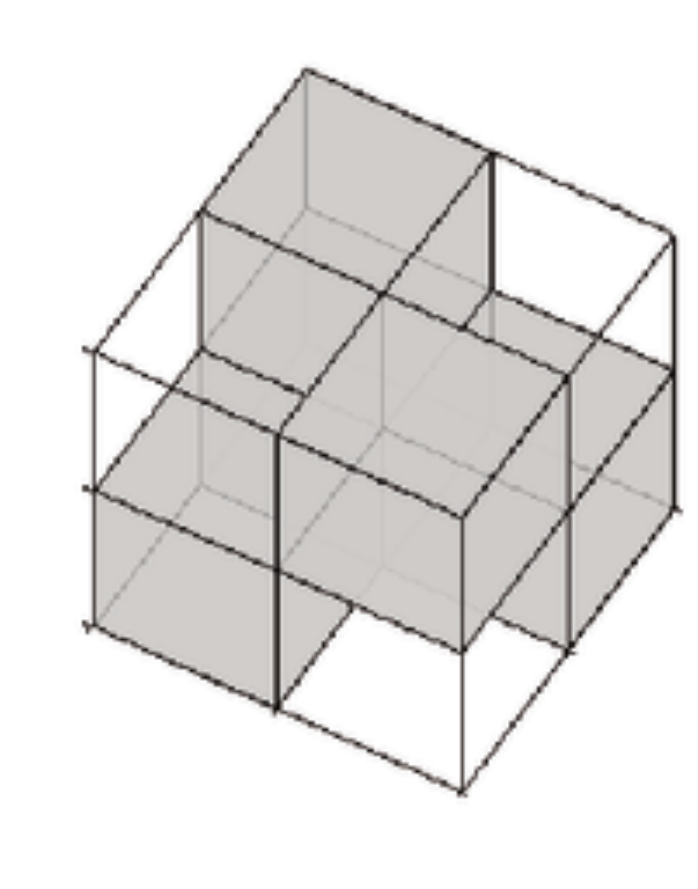}
  \mylab{-1.4cm}{-0.1cm}{$C_{53}$}
  \includegraphics[width=0.1\textwidth]{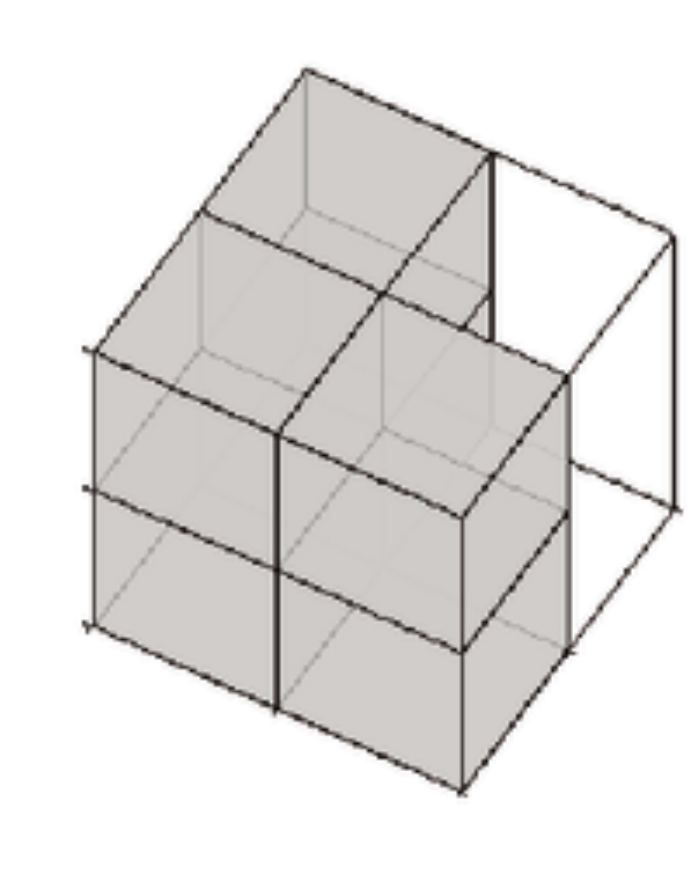}
  \mylab{-1.4cm}{-0.1cm}{$C_{61}$}
  \includegraphics[width=0.1\textwidth]{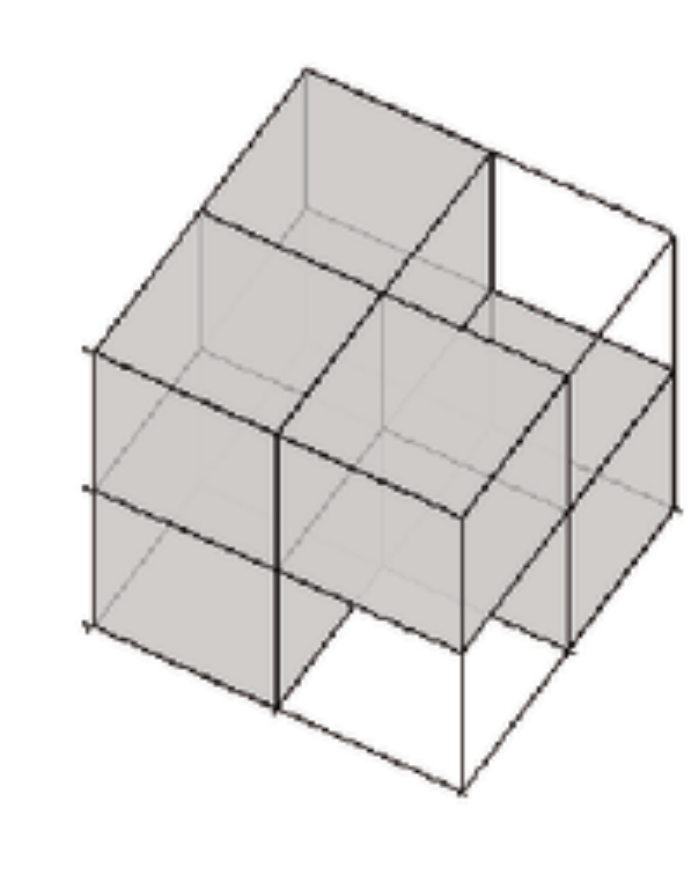}
  \mylab{-1.4cm}{-0.1cm}{$C_{62}$}
  \includegraphics[width=0.1\textwidth]{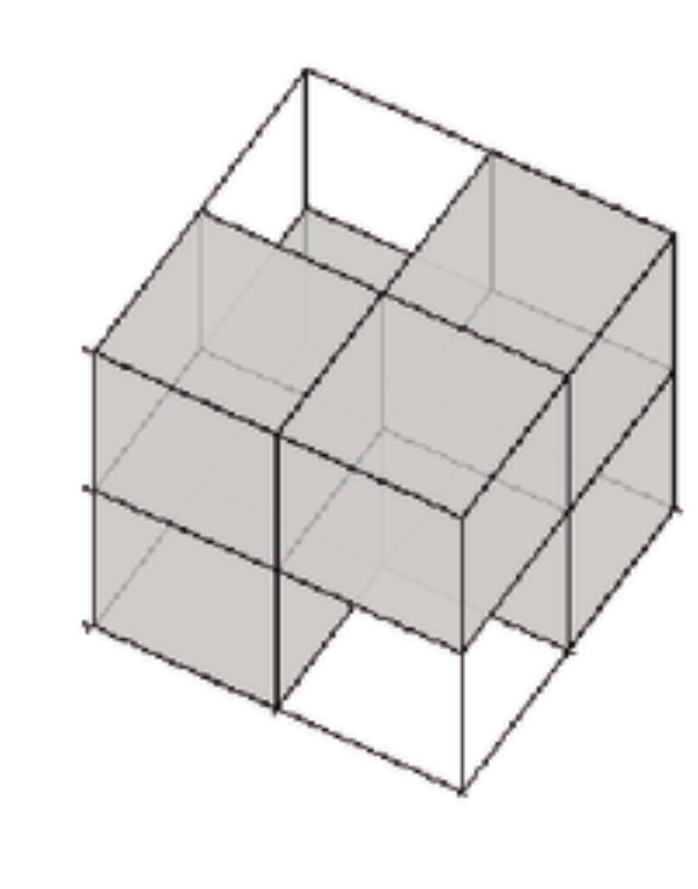}
  \mylab{-1.4cm}{-0.1cm}{$C_{63}$}
  \includegraphics[width=0.1\textwidth]{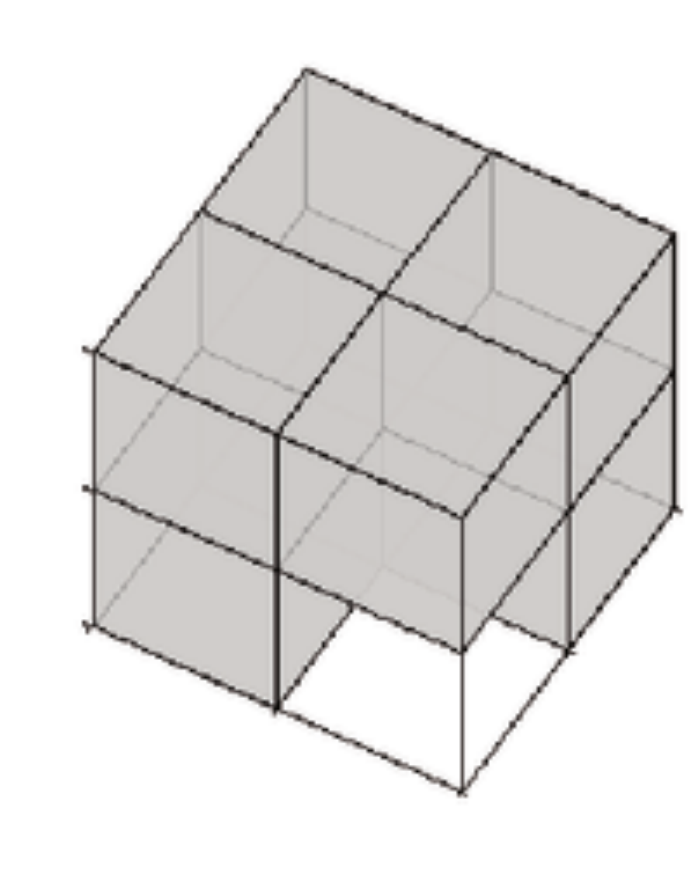}
  \mylab{-1.4cm}{-0.1cm}{$C_{71}$}
  \includegraphics[width=0.1\textwidth]{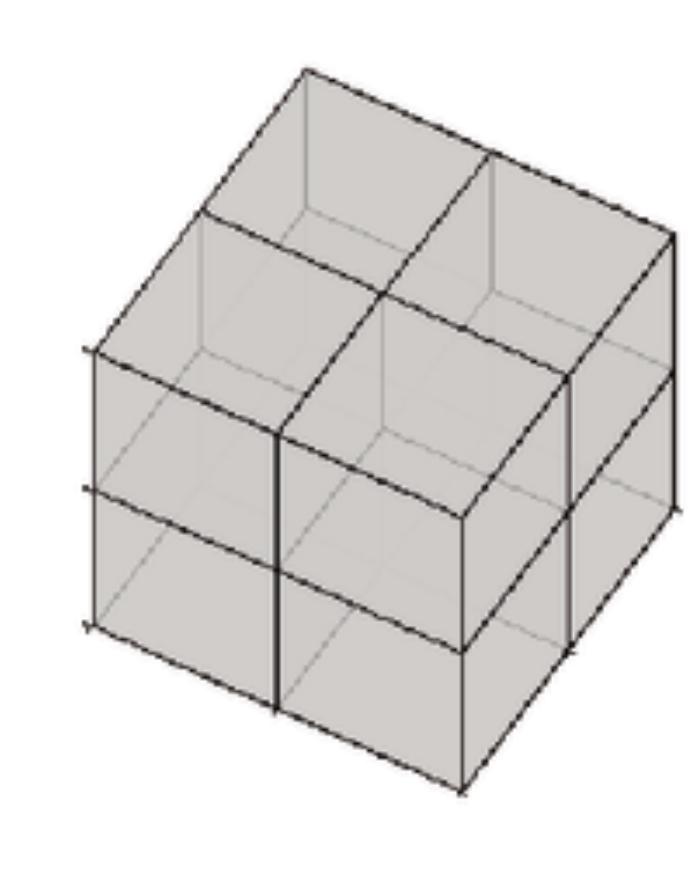}
  \mylab{-1.4cm}{-0.1cm}{$C_{81}$}
}
\caption{All possible configurations of voxels in a $2\times2\times2$
  sub-volume considering symmetries. Grey cubes represent voxels with
  value one. The cases are denoted by $C_{ij}$, where $i$ is the
  number of voxels with value one in the sub-volume.
\label{fig:aalgorithm_sketches}}
\end{figure}

The number of faces and edges at a particular vertex depends on its 8
surrounding voxels as shown in Figure \ref{fig:algorithm:cases}(b).
In this scenario, there are 256 different cases that may be reduced by
symmetry to those shown in Figure
\ref{fig:aalgorithm_sketches}. We will use the index
  $l$ to label sequentially the vertices of the virtual mesh. The
  contributions of the $l$-th vertex to the total number of faces,
  edges and vertices will be denoted by $\Delta F_l$, $\Delta E_l$ and
  $\Delta V_l$, respectively, and their values are tabulated in table
  \ref{table:aalgorithm} for all the possible cases. $F$, $E$ and $V$
  are then obtained as
\begin{eqnarray}\label{eq:FEV_aalgorithm}
F = \frac{\sum_l \Delta F_l}{4}, \quad E = \frac{\sum_l \Delta
  E_l}{2}, \quad V = \sum_l \Delta V_l,
\end{eqnarray}
where the summation extends to all the vertices of the virtual mesh.
Finally, the Euler characteristic and the genus are calculated with
(\ref{eq:euler_poly}) and (\ref{eq:euler}), respectively.
%
%_________________________________________________________________%
%_________________________________________________________________%
%
%----------------------------------------------------------------%
\begin{table}
  \tbl{Contribution to the number of faces $\Delta F$, edges $\Delta
    E$ and vertices $\Delta F$ of each configuration of voxels in the
    $2\times2\times2$ sub-domains shown in Figure
    \ref{fig:aalgorithm_sketches}.}  {
    \begin{tabular}{ccccccccccccc}
      Case     & $\Delta F$ & $\Delta E$ & $\Delta V$  & Case     & $\Delta F$ & $\Delta E$ & $\Delta V$  \\[1ex]
      \hline
      $C_{01}$ &     0     &      0      &      0      & $C_{44}$ &     8      &      8     &      2      \\
      $C_{11}$ &     3     &      3      &      1      & $C_{45}$ &     8      &      8     &      2      \\
      $C_{21}$ &     4     &      4      &      1      & $C_{46}$ &    12      &     12     &      4      \\
      $C_{22}$ &     6     &      6      &      2      & $C_{51}$ &     5      &      5     &      1      \\
      $C_{23}$ &     6     &      6      &      2      & $C_{52}$ &     7      &      7     &      1      \\
      $C_{31}$ &     5     &      5      &      1      & $C_{53}$ &     9      &      9     &      2      \\
      $C_{32}$ &     7     &      7      &      2      & $C_{61}$ &     4      &      4     &      1      \\
      $C_{33}$ &     9     &      9      &      3      & $C_{62}$ &     6      &      6     &      1      \\
      $C_{41}$ &     4     &      4      &      1      & $C_{63}$ &     6      &      6     &      2      \\
      $C_{42}$ &     6     &      6      &      1      & $C_{71}$ &     3      &      3     &      1      \\
      $C_{43}$ &     6     &      6      &      1      & $C_{81}$ &     0      &      0     &      0      \\
    \end{tabular}
      }
  \label{table:aalgorithm}
\end{table}
%----------------------------------------------------------------%

The alternative algorithm is now briefly described following the same
notation used in the previous section:
\begin{enumerate}
\item \emph{Initialize the variables}. $F, V$, and $E$ are equivalent to
  those described in section \ref{sec:algorithm} and are initialized
  to zero.
\item \emph{Loop through all the vertices in $A$}.  At the $l$-th
  vertex, proceed as follows:
  \begin{enumerate}
  \item \emph{Find the case}. Considering the 8 surrounding voxels at
    the $l$-th vertex shown in Figure \ref{fig:algorithm:cases}(b),
    search for the corresponding case in Figure
    \ref{fig:aalgorithm_sketches}, taking into account symmetries.
  \item \emph{Contributions to $F$, $E$ and $V$}. From table
    \ref{table:aalgorithm}, obtain $\Delta F_l$, $\Delta E_l$ and
    $\Delta V_l$ and compute $F\leftarrow F+\Delta F_l$, $E\leftarrow
    E+\Delta E_l$ and $V\leftarrow V+\Delta V_l$.
  \end{enumerate}
\item \emph{Compute the actual number of faces and edges}.
  $F\leftarrow F/4$ and $E\leftarrow E/2$.
\item \emph{Compute the Euler characteristic and genus}. $X = V - E +
  F$ and $G = (2 - X)/2$.
\end{enumerate}

The algorithm described in section \ref{sec:algorithm} is between 1.2
and 2 times faster than the one presented in this section, and roughly
3 times shorter in terms of lines of code, which makes it more
efficient and simple to implement.  For those reasons, the former
approach is preferred and the alternative algorithm is only considered
for validation purposes in the next section. \\

%==================================================================%
\section{Validation and scalability}\label{sec:validation}
%==================================================================%

Two approaches are followed to validate the algorithms detailed in
sections \ref{sec:algorithm} and \ref{sec:aalgorithm}. First,
synthetic cases whose genus are known beforehand are fed into the
algorithms, and the results are compared to the expected theoretical
values. Second, different datasets are used to compute the genus with
both algorithms and the outputs are shown to match.

\begin{figure}
\centerline{
  \includegraphics[width=0.40\textwidth]{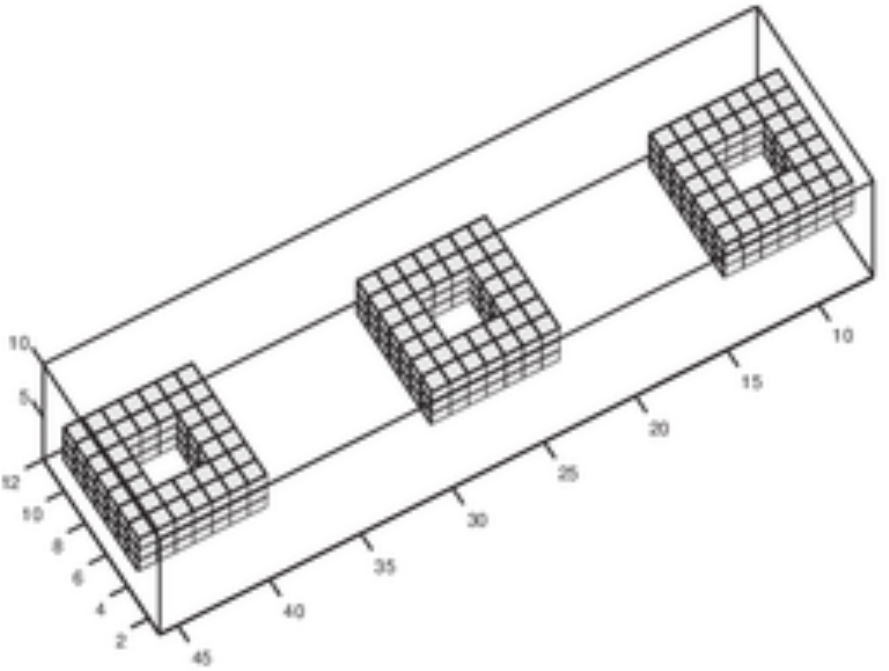}
  \mylab{-5.5cm}{3.2cm}{(\aaa)}
  \includegraphics[width=0.35\textwidth]{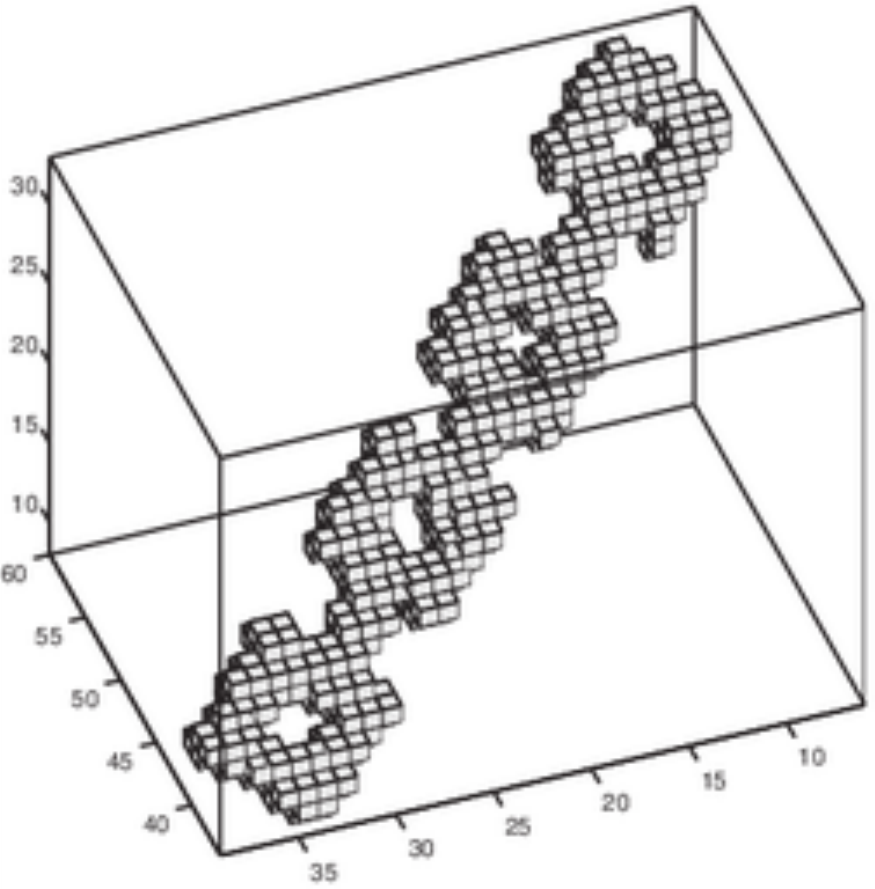}
  \mylab{-5.0cm}{4.5cm}{(\bbb)}
}
\centerline{
  \includegraphics[width=0.40\textwidth]{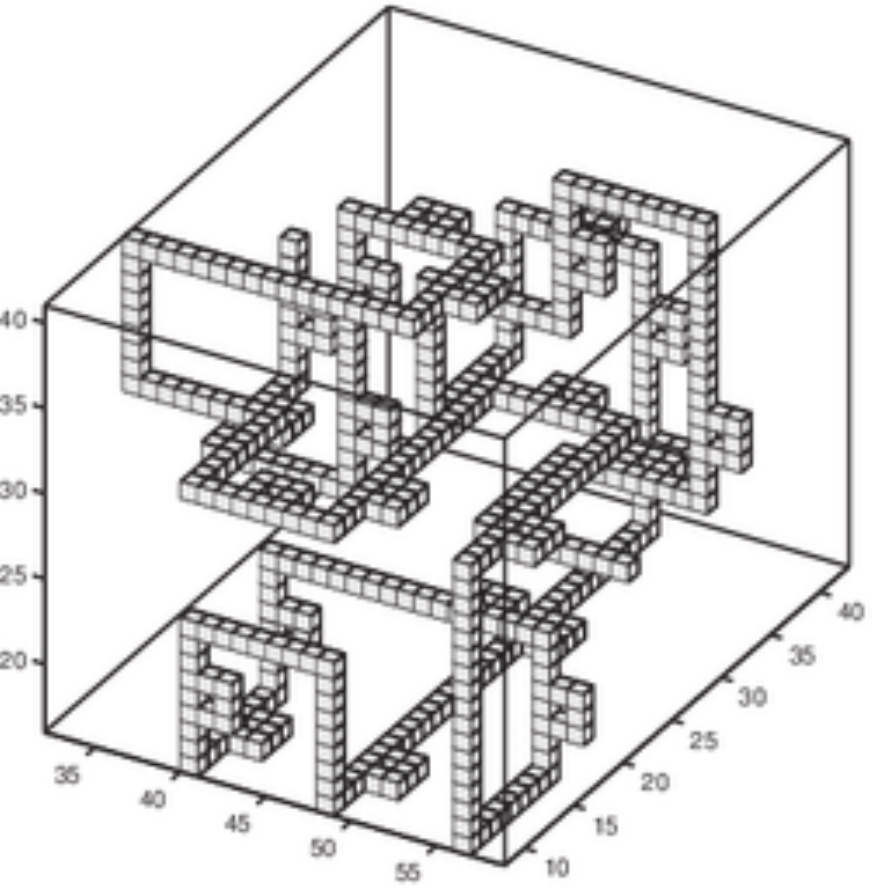}
  \mylab{-5.5cm}{5cm}{(\ccc)}
  \includegraphics[width=0.40\textwidth]{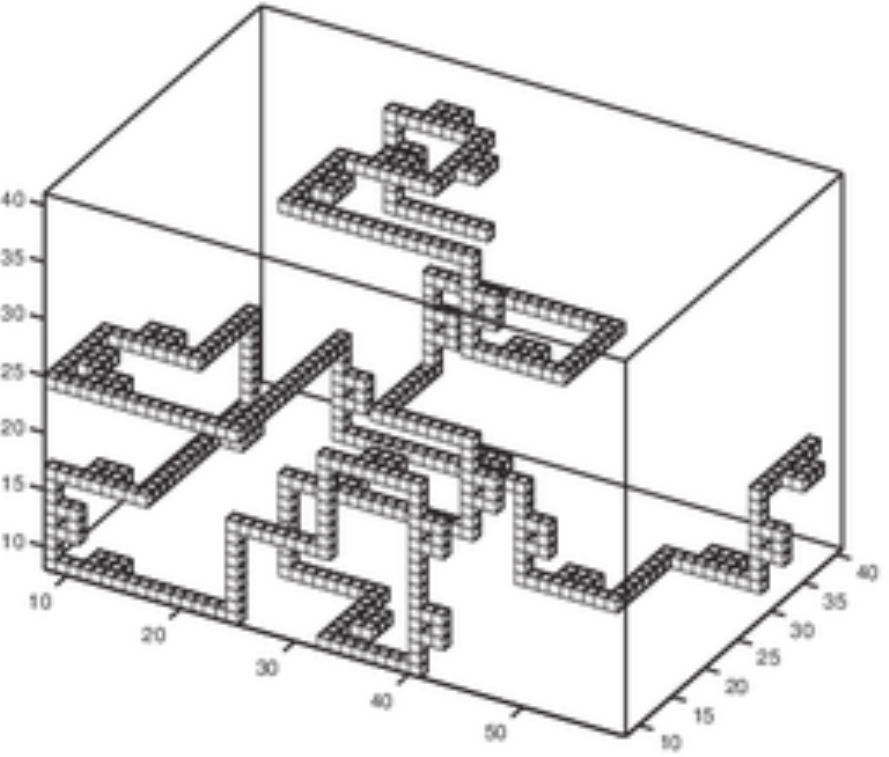}
  \mylab{-5.5cm}{5cm}{(\ddd)}
}
\caption{ Examples of synthetic cases used to validate the algorithms.
  (a), isolated torus; (b), concatenated torus; (c) and (d), synthetic
  random cases generated from the building blocks shown in Figure
  \ref{fig:validation:synthetic_pieces}.
\label{fig:validation:synthetic}}
\end{figure}

\begin{figure}
\centerline{
\raisebox{30pt}{\includegraphics[width=0.09\textwidth]{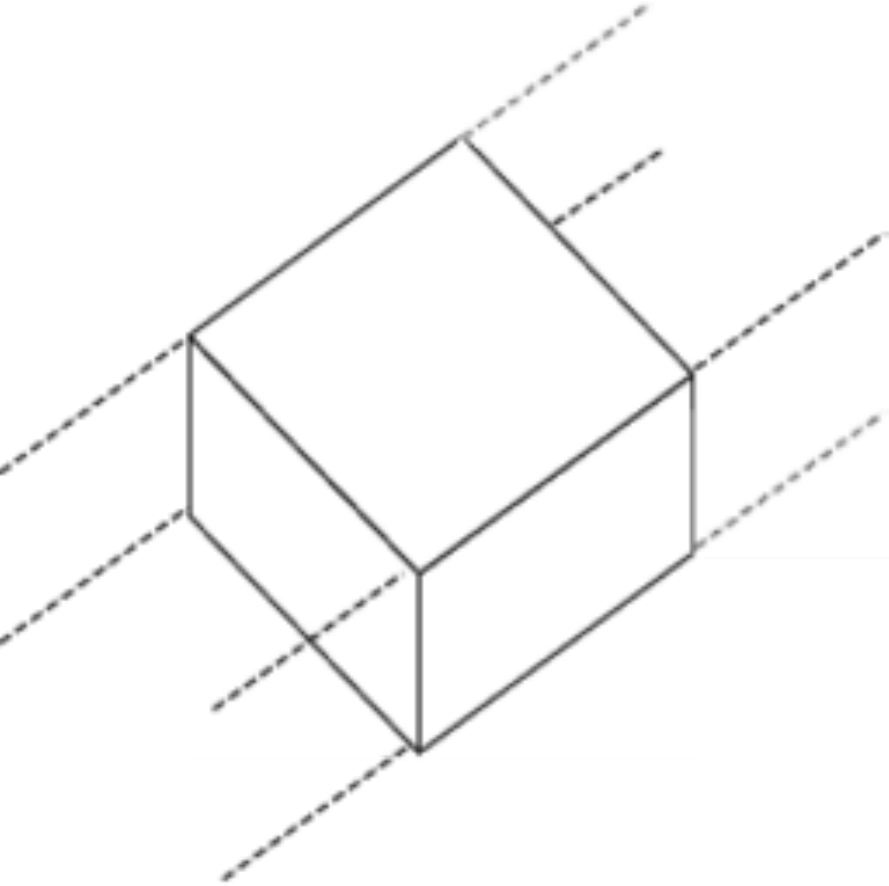}}
  \mylab{-0.9cm}{2.7cm}{(\aaa)}
  \hspace{0.5cm}
\raisebox{30pt}{\includegraphics[width=0.07\textwidth]{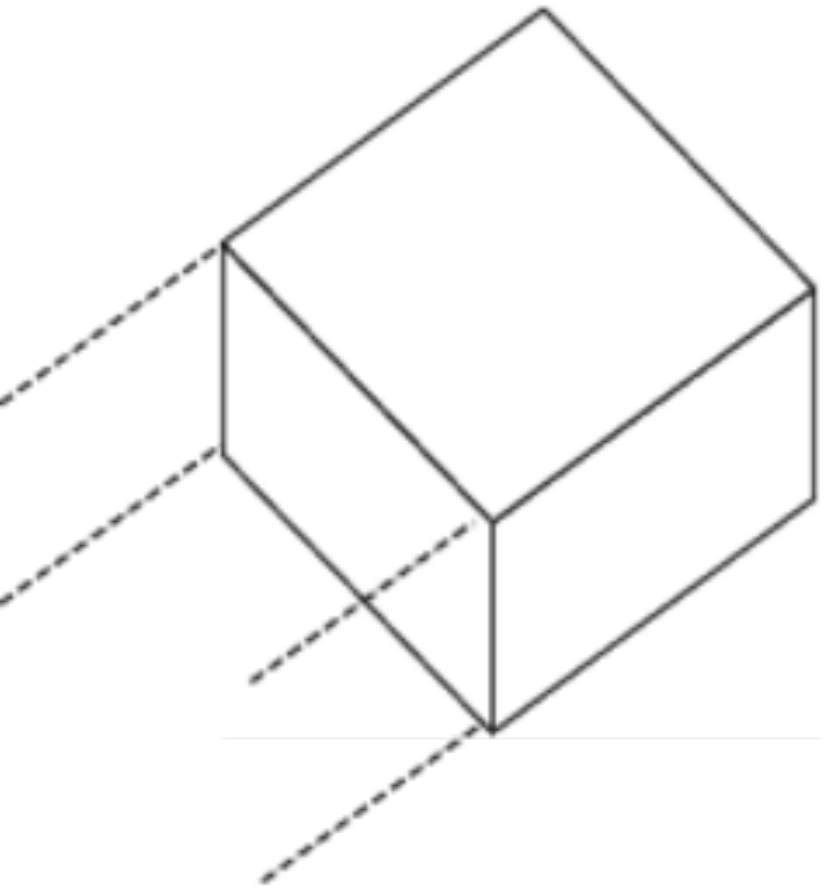}}
  \mylab{-0.7cm}{2.7cm}{(\bbb)}
  \hspace{0.5cm}
  \includegraphics[width=0.25\textwidth]{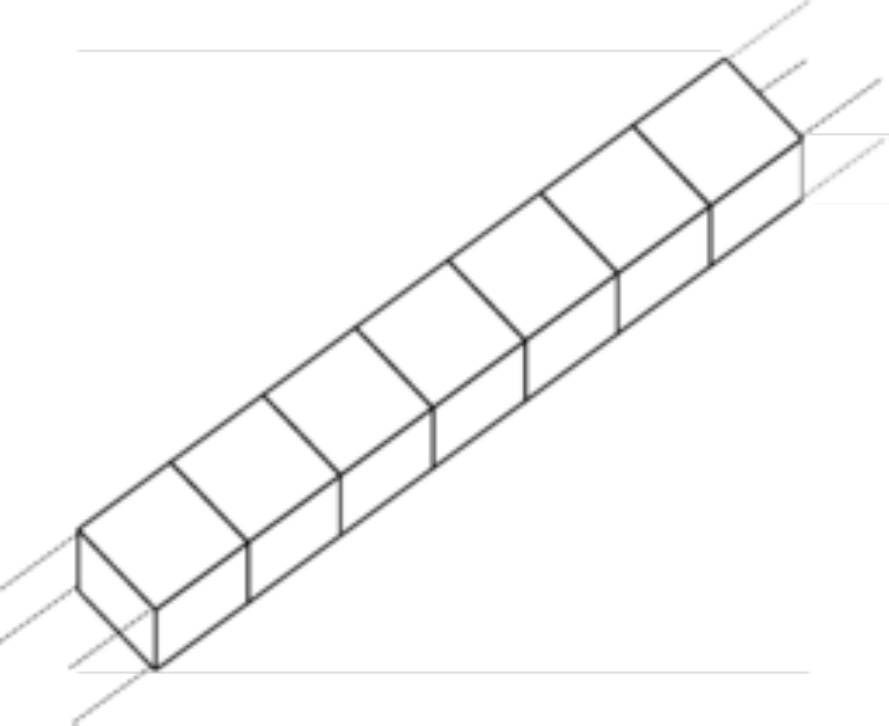}
  \mylab{-3cm}{2.7cm}{(\ccc)}
  \hspace{0.5cm}
  \includegraphics[width=0.25\textwidth]{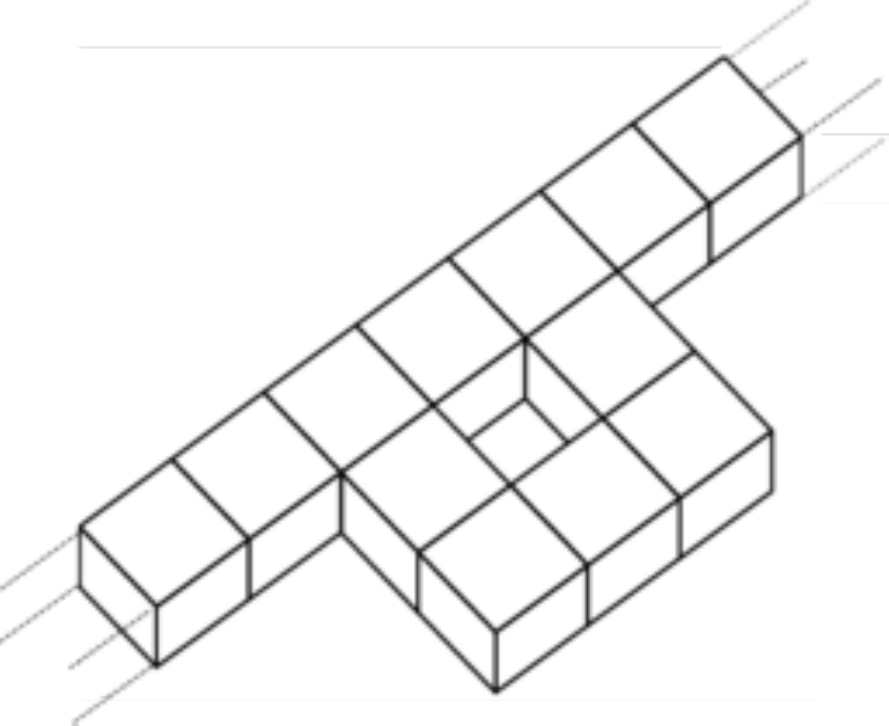}
  \mylab{-3cm}{2.7cm}{(\ddd)}
}
\caption{Building blocks of randomly generated test cases. (a), node;
  (b), end; (c), connector type I; (d), connector type II.
\label{fig:validation:synthetic_pieces}}
\end{figure}

% validation with synthetic data
% - all combination 2x2x2
% - isolated, donuts, donuts linked, random cases
The synthetic cases tested are the following: all possible
configurations in a $2\times2\times2$ volume, $n$ number of isolated
solid objects ($g=-n+1$), $n$ isolated objects with an interior cavity
each ($g=-2n+1$), $n$ isolated torus ($g=1$) as the example shown in
Figure \ref{fig:validation:synthetic}(a), and $n$ torus connected by
solid bridges ($g=n$) as in Figure
\ref{fig:validation:synthetic}(b). More cases were tested by rotating
the previous ones at different angles, for instance, as in the case
shown in Figure \ref{fig:validation:synthetic}(b). The values of $n$
tested range from $1$ to $10^6$. One more synthetic case tested
consists of randomly generated structures built using the blocks shown
in Figure \ref{fig:validation:synthetic_pieces}, referred to as
\emph{nodes} and \emph{ends} and linked by two type of
\emph{connectors}. The number of faces, edges and vertices of the
resulting object is given by
\begin{eqnarray}\label{eq:synthetic}
  F &=& n_e \Delta F_e + n_n \Delta F_n + n_I \Delta F_I + n_{II}
  \Delta F_{II}, \\ E &=& n_e \Delta E_e + n_n \Delta E_n + n_I \Delta
  E_I + n_{II} \Delta E_{II}, \\ V &=& n_e \Delta V_e + n_n \Delta V_n
  + n_I \Delta V_I + n_{II} \Delta V_{II},
\end{eqnarray}
where $n_e$, $n_n$, $n_I$ and $n_{II}$ are the number of ends, nodes,
and connectors of type I and II that belong to the object. The
increments $\Delta F_i$, $\Delta E_i$ and $\Delta V_i$ with $i=e,b,I$
and $II$ are the contribution to the number of faces, edges and
vertices of each block respectively, and its values are tabulated in
table \ref{table:synthetic}. Roughly $10^6$ cases were randomly
generated and tested and two examples are shown in Figures
\ref{fig:validation:synthetic}(c,d). More synthetic cases similar to
those presented above but using differently shaped connectors were
also successfully tested (not shown).
%
%_________________________________________________________________%
%_________________________________________________________________%
%
%_________________________________________________________________%
%_________________________________________________________________%
%
%----------------------------------------------------------------%
\begin{table}
  \tbl{Contribution to the number of faces $\Delta F$, edges
    $\Delta E$ and vertices $\Delta V$ of the different blocks shown
    in Figure \ref{fig:validation:synthetic_pieces}.}{
    \begin{tabular}{lccccc}
       Case      & end   & node & connector type I &  connector type II \\[1ex]
      \hline
      $\Delta F$ &   5   &  4   &      28      &        46    \\
      $\Delta E$ &  12   & 12   &      52      &        88    \\
      $\Delta V$ &   8   &  8   &      24      &        40     \\
    \end{tabular}}
  \label{table:synthetic}
\end{table}
%----------------------------------------------------------------%

\begin{figure}
\centerline{ 
  \includegraphics[width=0.30\textwidth]{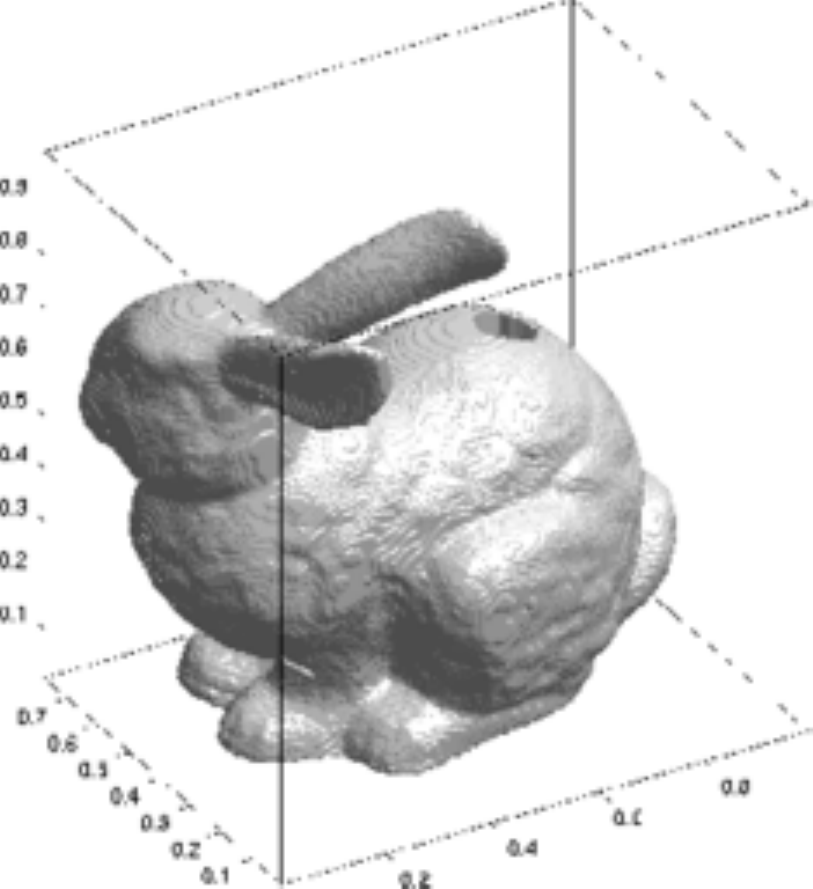}
  \mylab{-4cm}{4.5cm}{(\aaa)}
  \hspace{0.2cm}
  \includegraphics[width=0.20\textwidth]{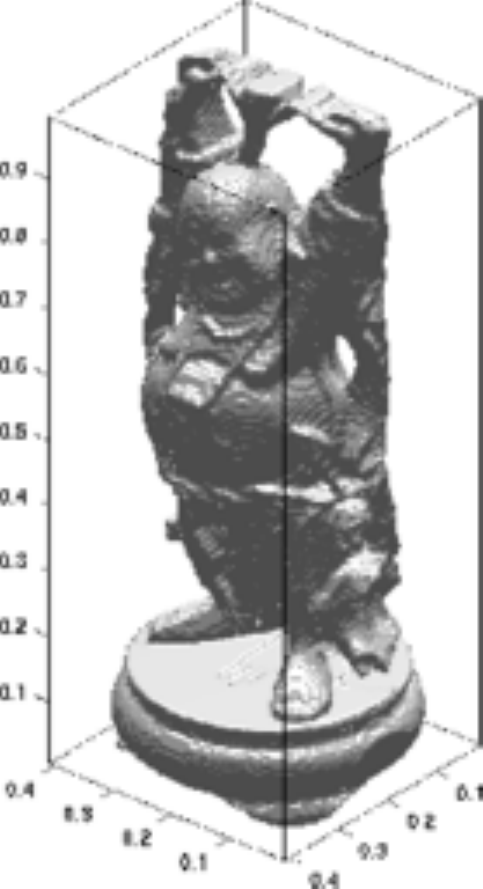}
  \mylab{-3cm}{5cm}{(\bbb)}
  \hspace{0.2cm}
  \includegraphics[width=0.30\textwidth]{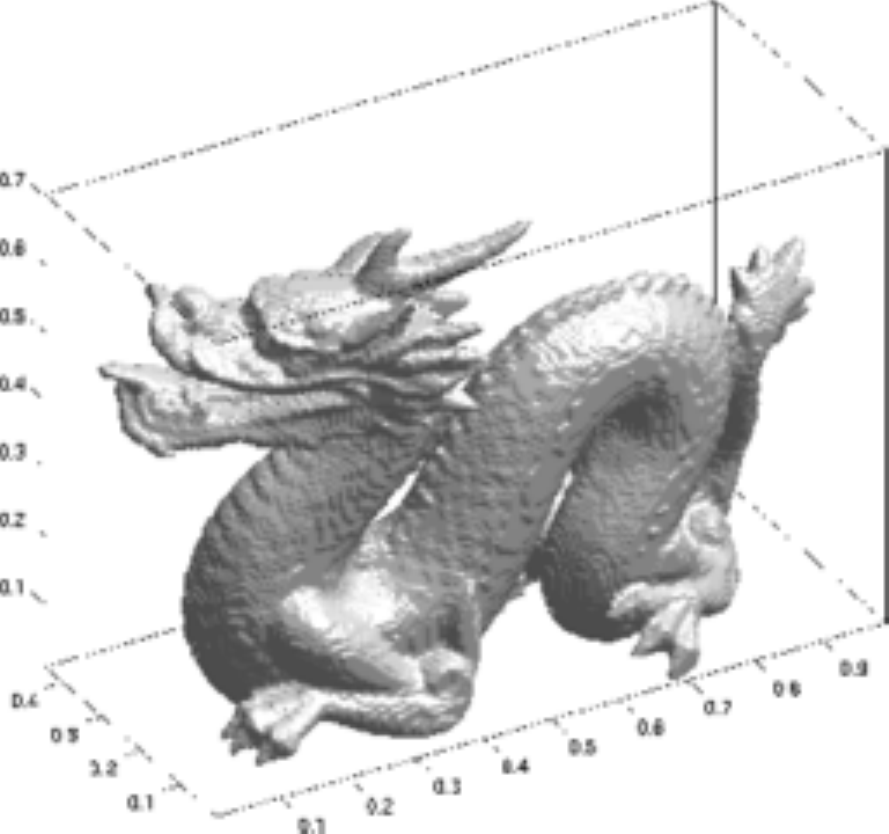}
  \mylab{-3.5cm}{4.5cm}{(\ccc)}
}
\caption{3D models obtained from the Stanford 3D Scanning Repository
  and voxelized with \emph{binvox}. (a), bunny \cite{Turk:1994}; (b),
  Buda \cite{Curless:1996}; (c), dragon \cite{Curless:1996}.
\label{fig:validation:compare_models}}
\end{figure}

\begin{figure}
\centerline{
  \includegraphics[width=0.45\textwidth]{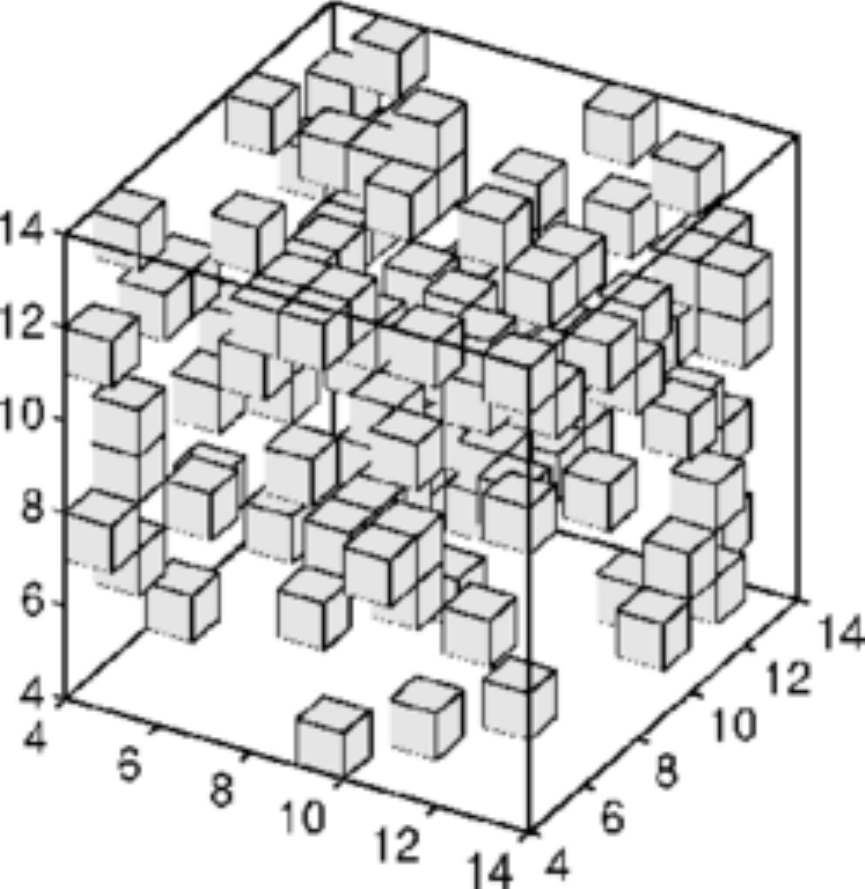}
  \mylab{-6cm}{6cm}{(\aaa)}
  \hspace{0.2cm}
  \includegraphics[width=0.45\textwidth]{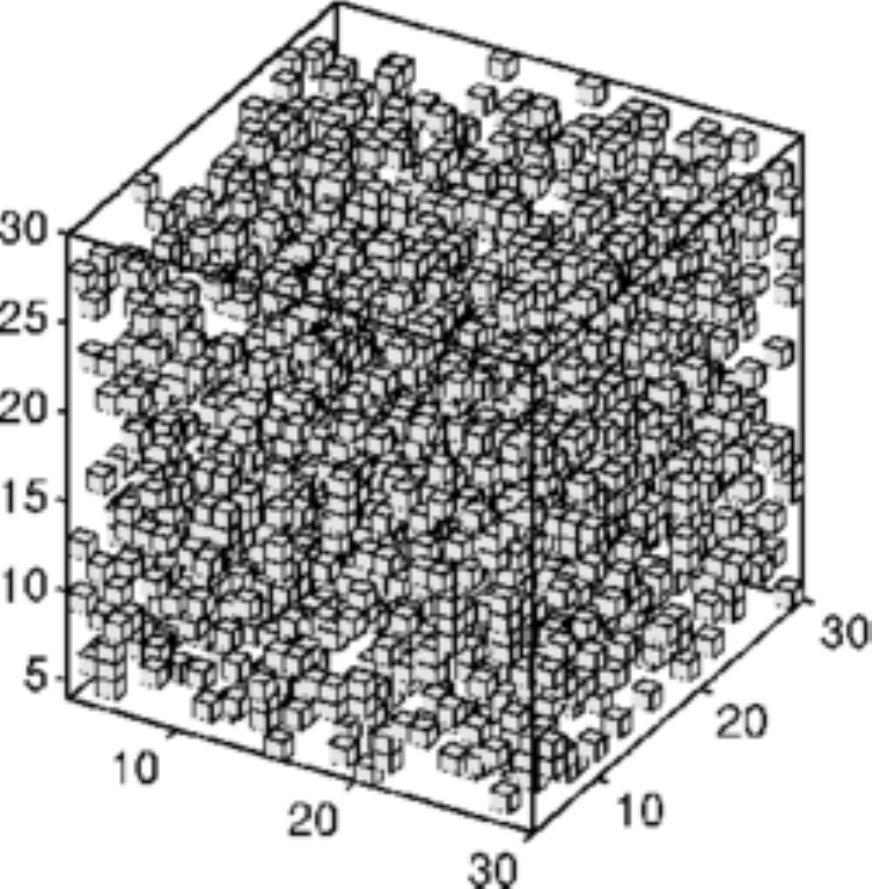}
  \mylab{-6cm}{6cm}{(\bbb)}
}
\caption{Test cases of cubical domains with (a), $16^3$ and (b),
  $32^3$ voxels randomly initialized with 0's and 1's.
\label{fig:validation:compare_random}}
\end{figure}

% validation comparing methods
% - 3D models
% - random data
We perform a second validation comparing the number of faces, edges
and vertices computed with the algorithm presented in section
\ref{sec:algorithm} and the alternative one in section
\ref{sec:aalgorithm}, which of course, must be identical. This was
verified for the synthetic cases described above.  More test cases are
the three models from the Stanford 3D Scanning Repository
\cite{stanford} voxelized with \emph{binvox} \cite{binvox} (see also
\citeN{noo:tur:2003}) and shown in Figure
\ref{fig:validation:compare_models}. Finally, we tested $10^2$ cases
delimited by a cubical region and with grid sizes from $16^3$ up to
$4096^3$ whose voxels were randomly initialized with 0's and 1's
filling approximately 50\% of the total volume. Two examples are shown
in Figure \ref{fig:validation:compare_random}. The two algorithms
yield identical results for all the cases tested, counting exactly the
same number of faces, edges and vertices, and therefore, the same
genus.

Table \ref{table:validations} summarizes the number of voxels, faces,
edges, vertices and genus of some of the cases tested, which are
available for download in our webpage \cite{torroja}.
%
%_________________________________________________________________%
%_________________________________________________________________%
%
%_________________________________________________________________%
%_________________________________________________________________%
%
%
%----------------------------------------------------------------%
\begin{table}
  \tbl{Summary of some of the datasets tested and available for
    download at \cite{torroja} }{
    %    \begin{tabular}{lccccc}
    \begin{tabular}{lcrrrr}
      Case  & Size & Faces & Edges & Vertices & Genus  \\[1ex]
      \hline
      Synthetic1  & $64^3$   &         1924 &         3848 &        1880 &         23 \\
      Synthetic2  & $64^3$   &         2174 &         4348 &        2120 &         28  \\
      Bunny       & $256^3$  &       309482 &       618964 &      309466 &          9 \\
      Buda        & $256^3$  &       129800 &       259600 &      129780 &         11 \\
      Dragon      & $256^3$  &       164494 &       328988 &      164494 &          1 \\
      Random1     & $64^3$   &       297496 &       594992 &      280160 &       8669 \\
      Random2     & $128^3$  &      2744830 &      5489660 &     2570182 &      87325 \\ 
      Random3     & $256^3$  &     23530742 &     47061484 &    21985520 &     772612  \\ 
      Random4     & $512^3$  &    194709102 &    389418204 &   181726644 &    6491230 \\ 
      Random5     & $1024^3$ &   1584014008 &   3168028016 &  1477589086 &   53212462  \\ 
      Random6     & $2048^3$ &  12778133206 &  25556266412 & 11916193918 &  430969645  \\ 
      Random7     & $4096^3$ & 102651228492 & 205302456984 & 95713851166 & 3468688664  \\ 
    \end{tabular}}
  \label{table:validations}
\end{table}
%----------------------------------------------------------------%

\begin{figure}
\centerline{
   \psfrag{X}{ \raisebox{-7pt}{$N$} }
   \psfrag{Y}[bc][bl]{ \raisebox{-15pt}{$time$ (s) } }
   \psfrag{a}[][][0.7]{ \raisebox{-5pt}{ $64^3$} }
   \psfrag{b}[][][0.7]{ \raisebox{-5pt}{ $128^3$} }
   \psfrag{c}[][][0.7]{ \raisebox{-5pt}{ $256^3$} }
   \psfrag{d}[][][0.7]{ \raisebox{-5pt}{ $512^3$} }
   \psfrag{e}[][][0.7]{ \raisebox{-5pt}{ $1024^3$} }
   \psfrag{f}[][][0.7]{ \raisebox{-5pt}{ $2048^3$} }
   \psfrag{g}[][][0.7]{ \raisebox{-3pt}{ $4096^3$} }
   \includegraphics[width=0.47\textwidth]{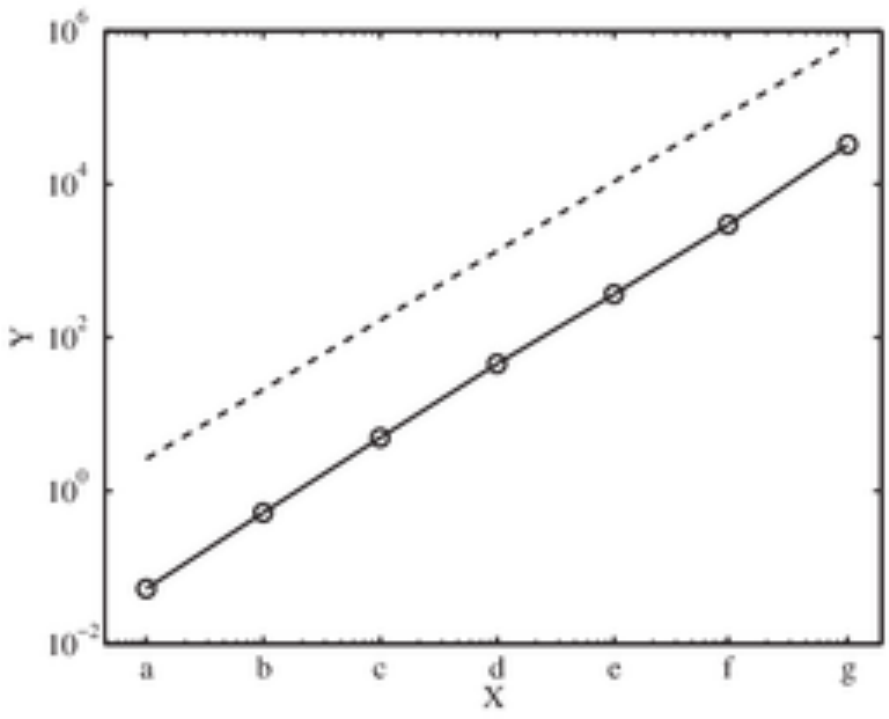}
   \mylab{-5.5cm}{4.5cm}{(\aaa)}
   \hspace{0.5cm}
   \psfrag{X}{ \raisebox{-4pt}{$n_{proc}$} }
   \psfrag{Y}[bc][bl]{ \raisebox{0pt}{Efficiency} }
   \includegraphics[width=0.468\textwidth]{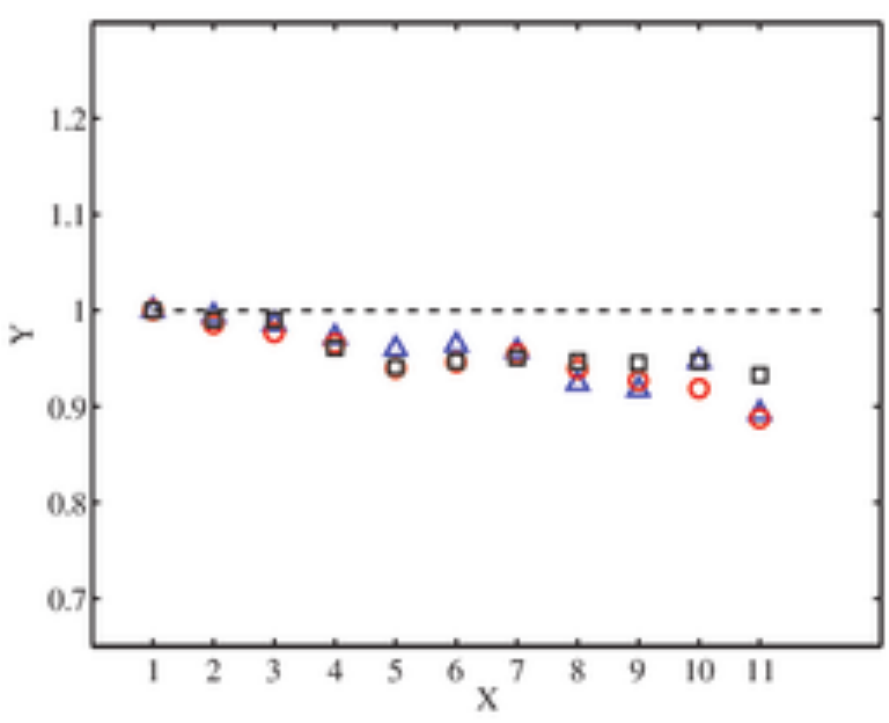}
   \mylab{-5.5cm}{4.5cm}{(\bbb)}
}
\caption{ (a) For the Fortran serial version, average time in seconds
  elapsed to compute the genus of cubical arrays of size $N^3$
  randomly initialized to 0's and 1's with roughly 50\% of the volume
  occupied. Results for a single processor. The circles are the
  measured times and the solid dashed line is $time \sim N$. (b) For
  the Fortran Coarrays version, strong scaling efficiency as a
  function of the number of processing elements, $n_{proc}$, for three
  different problem sizes, $N=1024^3$ (\trian), $N=2048^3$ (\circle),
  and $N=4096^3$ (\squar).
\label{fig:validation:scalability}}
\end{figure}

% scalability
The algorithm presented in section \ref{sec:algorithm} was implemented
in Fortran, compiled with Intel Fortran Studio XE 2016 16.0.0 20150815
, and tested in an Intel\textsuperscript{\textregistered}
Xeon\textsuperscript{\textregistered} CPU X5650 2.67GHz with 192GiB of
RAM for cubical arrays of different sizes and randomly generated as
those shown in table \ref{table:validations}.  The average time
elapsed to compute the genus of inputs with different sizes is
presented in Figure \ref{fig:validation:scalability}(a), which shows
linear scalability and makes feasible applications to very large
datasets.

The code was also parallelized using Fortran Coarrays. The domain
decomposition was performed by dividing the $z$ direction in chunks of
size $N_x \times N_y \times \Delta N_z$, where $\Delta N_z$ is
nint$(N_z/n_{proc})$, and $n_{proc}$ the number of processing
elements. Two overlapped $x$-$y$ planes are added at the beginning and
end of each chunk in order to compute faces, edges and vertices
without any extra communication between images. Once this is done, the
genus is obtained by summing the faces, edges and vertices of all the
chunks. The parallelization works for any number of processing
elements smaller than $N_z$, and the size of the last chunk may differ
from the size of the others if $N_z$ is not divisible by
$n_{proc}$. The reader is referred to the software component of the
manuscript to cover all the details of the parallelization. The strong
scaling efficiency, where the problem size stays fixed but the number
of processing elements increases, is shown in Figure
\ref{fig:validation:scalability}(b). The results are quite
satisfactory and the efficiency remains always above 90\%.
%
%_________________________________________________________________%
\\
%_________________________________________________________________%

%==================================================================%
\section{Applications to turbulent flows}\label{sec:applications}
%==================================================================%

We show two examples where the genus is used as a tool to characterize
the topology of regions of interest in turbulent flows. In the first
example, the genus is computed for millions of individual coherent
structures extracted from a turbulent channel flow. In the second one,
the genus is used to identify physically meaningful interfaces
separating turbulent and non-turbulent flow in a time-decaying jet.

%-----------------------------------------------------------------%
\subsection{Topology of coherent regions in turbulent flows}
\label{subsec:applications:coherent}
%-----------------------------------------------------------------%
  
%_________________________________________________________________%
\begin{figure}
\centerline{
%\mylab{1cm}{3.5cm}{(\aaa)}
\includegraphics[width=0.32\textwidth]{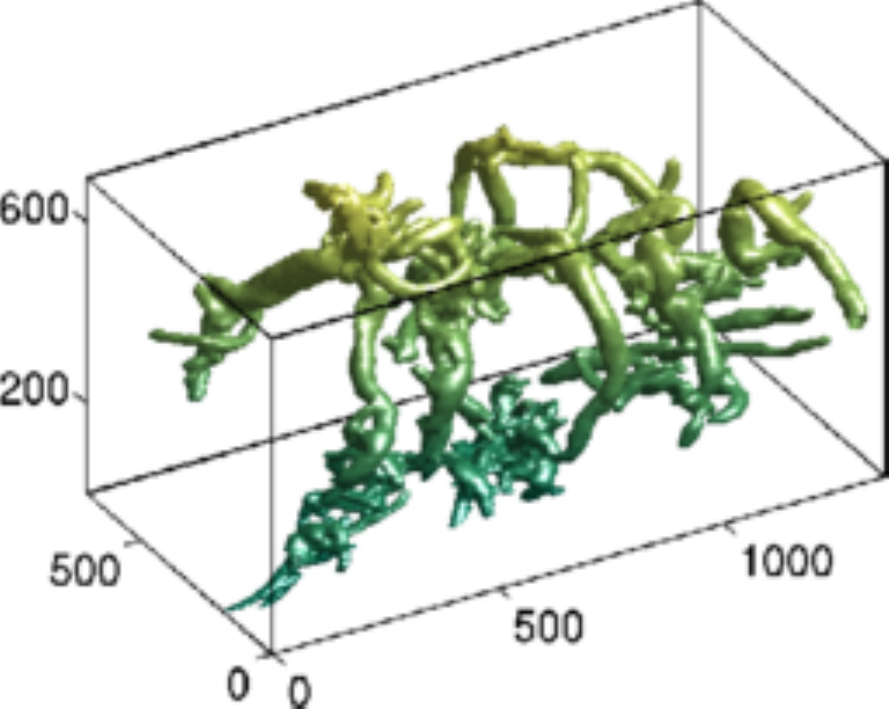}
%\mylab{1.3cm}{3.5cm}{(\bbb)}
\includegraphics[width=0.32\textwidth]{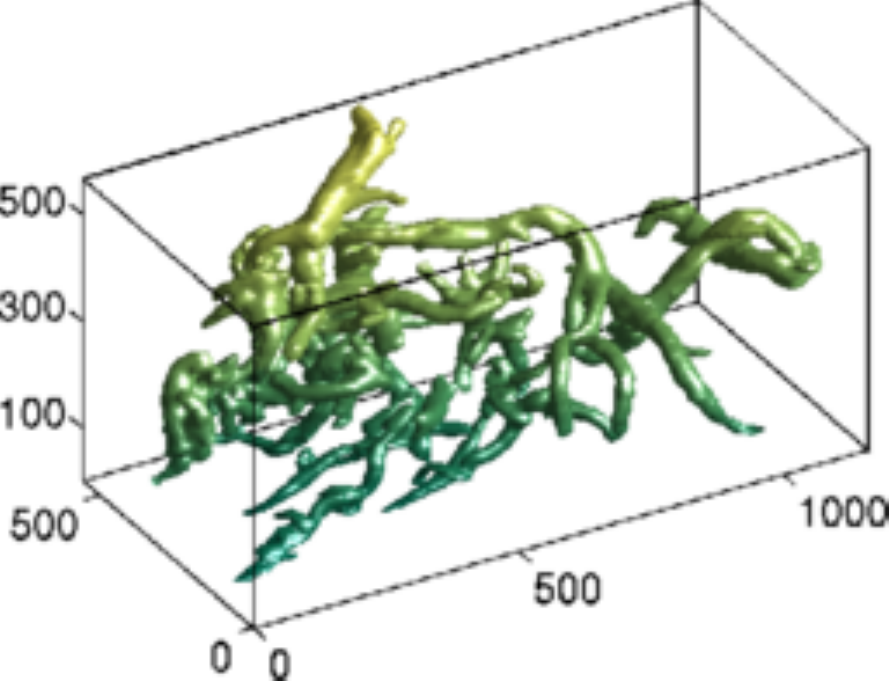}
%\mylab{1.3cm}{3.5cm}{(\ccc)}
\includegraphics[width=0.32\textwidth]{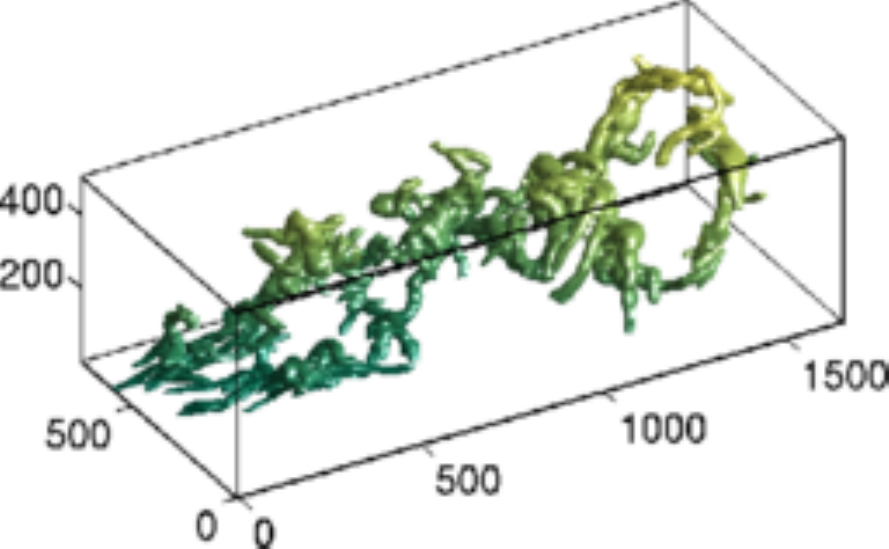}
}
\centerline{
%\mylab{1cm}{3.2cm}{(\ddd)}
\includegraphics[width=0.42\textwidth]{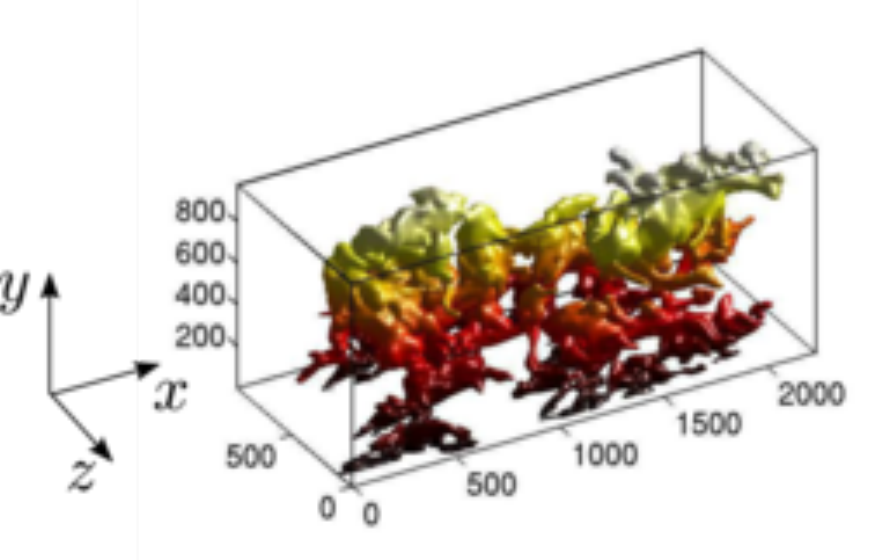}
%\mylab{1.3cm}{3.2cm}{(\eee)}
\includegraphics[width=0.22\textwidth]{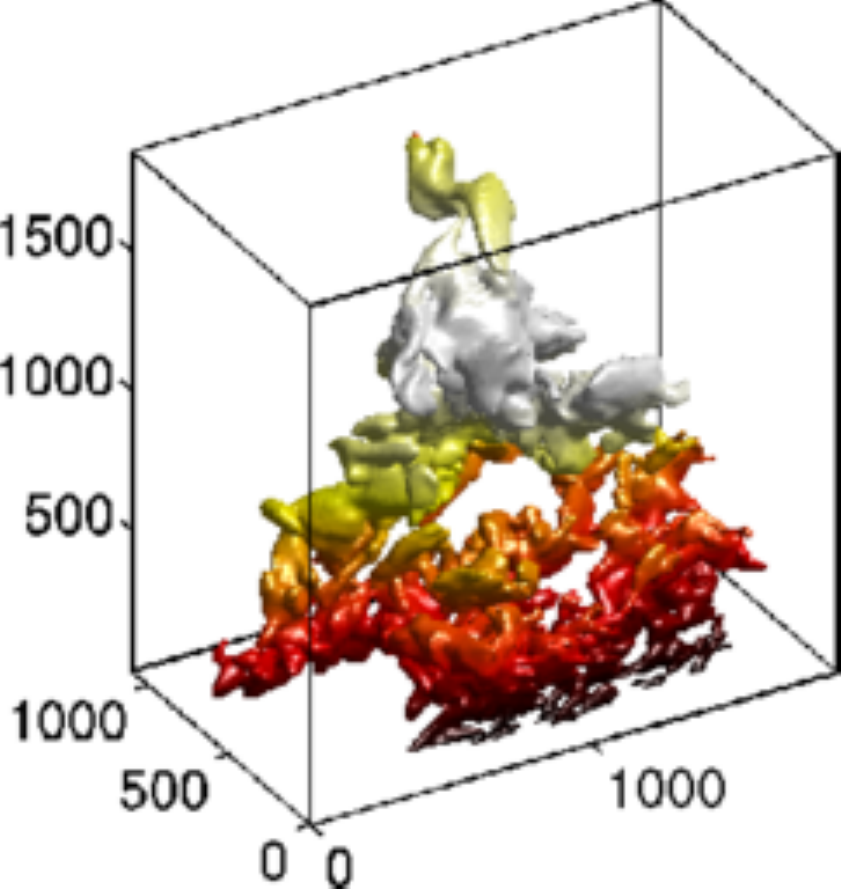}
%\mylab{1.3cm}{3.2cm}{(\fff)}
\includegraphics[width=0.33\textwidth]{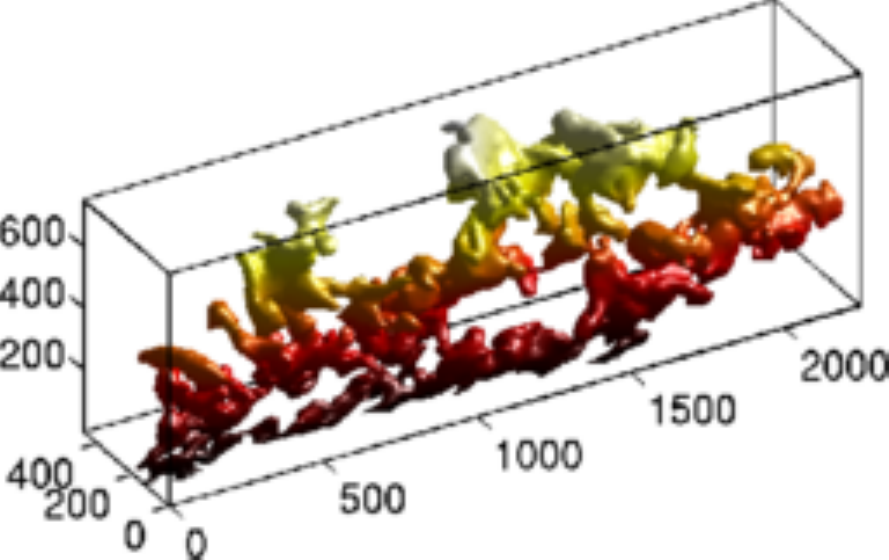}
}
\caption{Examples of three vortex cluster (top row) and Q-structures
  (bottom row) extracted from a direct numerical simulation of a
  turbulent channel at $Re_\tau=4180$ \cite{loz:jim:2014}. The flow
  goes from bottom-left to top-right. The axis are normalized with
  $\nu/u_\tau$. The colors change gradually with the distance to the
  wall, which is located at $y=0$. Note that the objects are not to
  scale.
\label{fig:applications:coherent}}
\end{figure}
%_________________________________________________________________%

We use three direct numerical simulations of turbulent channel flows
(two parallel walls delimiting a flow moving on average in one
direction) from \citeN{loz:jim:2014} at Reynolds numbers
$Re_\tau=934,2004$ and $4180$, with $Re_\tau= h u_\tau/\nu$ where $h$
is the channel half-height, $u_\tau$ the friction velocity and $\nu$
the kinematic viscosity. More details about turbulent channel flows
may be found in \citeN[Chapter~7.1]{pope:2000}. The streamwise,
wall-normal and spanwise directions are denoted by $x$, $y$ and $z$
respectively.  Very briefly, we compute the genus of coherent
structures, namely, regions of the flow where a variable is higher
than a prescribed threshold.  The three-dimensional coherent
structures under study are vortex cluster from
\citeN{ala:jim:zan:mos:2006} and Q-structures from
\citeN{loz:flo:jim:2012}.  The former are defined in terms of the
discriminant of the velocity gradient and are connected regions
satisfying
\begin{equation}\label{eq:thr:clus}
D(x,y,z)/D'(y)>\alpha,
\end{equation}
where $D$ is the instantaneous discriminant of the velocity gradient
tensor, $D'(y)$ its standard deviation at each $x-z$ plane and $\alpha
= 0.02$ a thresholding parameter obtained from a percolation
analysis. Similarly, Q-structures are defined as places where
\begin{equation}\label{eq:thr:uvs}
uv(x,y,z)/uv'(y)>H,
\end{equation}
where $uv$ is the instantaneous tangential Reynolds stress, being $u$
and $v$ the streamwise and wall-normal velocity fluctuations, $uv'(y)$
its rooted-mean-squared value at each $y$-position, and $H$ a
thresholding parameter equal to $1.75$.  Three-dimensional objects are
constructed by connecting neighboring grid points fulfilling relations
(\ref{eq:thr:clus}) for vortex clusters and (\ref{eq:thr:uvs}) for
Q-structures and using the 6-connectivity criteria.  Full details for
both types of structures can be found in \citeN{ala:jim:zan:mos:2006},
\citeN{loz:flo:jim:2012} and \citeN{loz:jim:2014}. To compute the
genus, each object is circumscribed within a box aligned to the
Cartesian axes which constitutes the limits of the array $A(i,j,k)$
discussed in section \ref{sec:algorithm}. Figure
\ref{fig:applications:coherent} shows several examples of actual
objects extracted from the flow and demonstrates the complex
geometries that may appear. The number of structures computed is of
the order of $10^7$, with a wide spectrum of sizes ranging from
$\sim30^3$ to $\sim 2000^3$ voxels.
%
%_________________________________________________________________%
\begin{figure}
\centerline{
\psfrag{X}{$g$}
\psfrag{Y}{PDFs}
\includegraphics[width=0.48\textwidth]{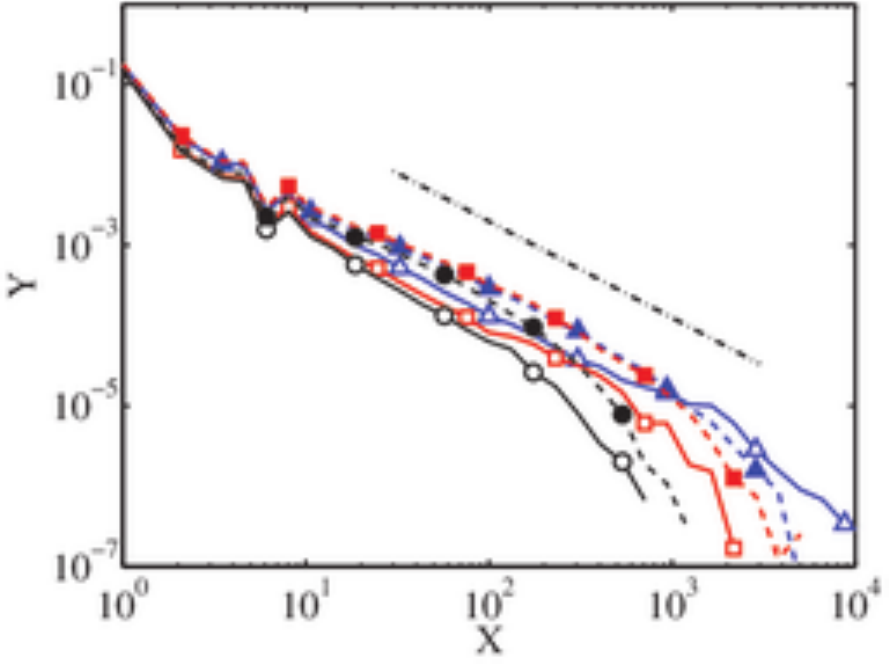}
\mylab{-1cm}{4.5cm}{(\aaa)}
\psfrag{X}{ \raisebox{-2pt}{$V_o/\eta^3$} }
\psfrag{Y}{$\langle g \rangle$}
\includegraphics[width=0.48\textwidth]{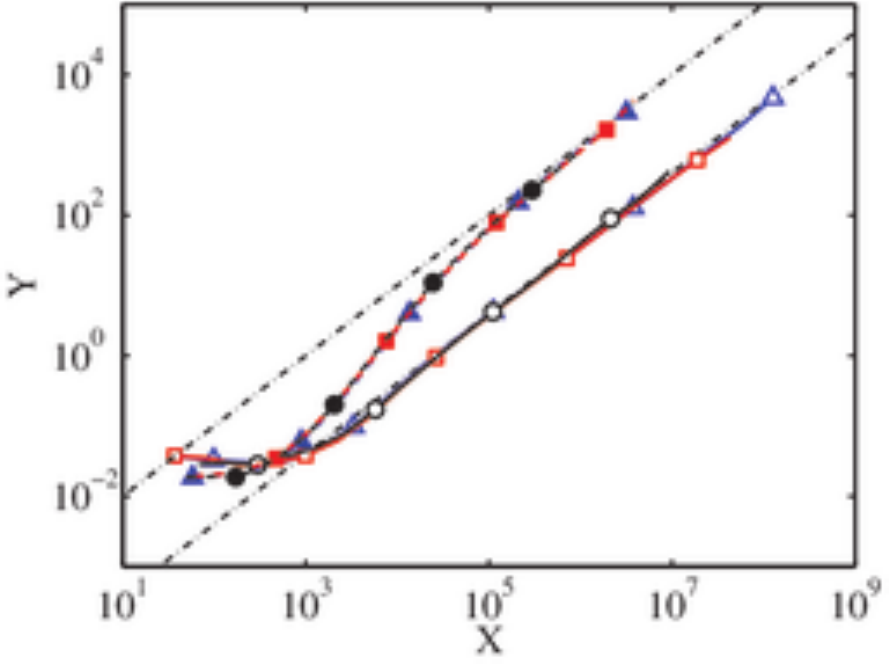}
\mylab{-5.5cm}{4.5cm}{(\bbb)}
}
\caption{ (a) Probability density functions of the genus. The
  dashed-dotted line is proportional to $g^{-1.2}$.  (b) Average
  number of holes (genus) of individual coherent structures as a
  function of their volume, $V_o$, in Kolmogorov units. The
  dashed-dotted lines are $\langle g \rangle
  =10^{-3}\eta^{-3}(V_o/\eta^3)$ and $\langle g \rangle=4\times
  10^{-5}\eta^{-3}(V_o/\eta^3)$.  For (a) and (b), the solid lines
  with open symbols correspond to Q-structures and the dashed lines
  with closed symbols to vortex clusters.  Different symbols stand for
  different Reynolds numbers; \circle, $Re_\tau=934$; \squar,
  $Re_\tau=2004$; \trian, $Re_\tau=4180$.
\label{fig:applications:genus_clusters}}
\end{figure}
%_________________________________________________________________%

Each array $A(i,j,k)$ contains just one single object and, hence, the
only contributions to the genus are the number of holes and internal
cavities. The data reveals that only 0.05\% of objects have negative
genus and it was checked that most of structures are solid.  In this
scenario, genus and number of holes can be used interchangeably.

The probability density functions (PDFs) of the genus, $g$, are
presented in Figure \ref{fig:applications:genus_clusters}(a) and most
of the values concentrate around zero or a few holes, although the
long potential tails reach values up to $10^4$ holes.  Figure
\ref{fig:applications:genus_clusters}(b) shows the average number of
holes in the objects as a function of their volume, $V_o$, normalized
in Kolmogorov units, $\eta^3$ (see \citeN[Chapter~6]{pope:2000}).  It
becomes clear that as the volume of the structures increases, so does
the genus, which is reasonable if we consider that the volume of the
object is related to its internal Reynolds number (or complexity), and
increasing its volume results in more complicated topologies. The
curves for both vortex clusters and Q-structures show good collapse
for the three Reynolds numbers and follow the trend $\langle g \rangle
= \rho V_o$, with $\langle g \rangle$ the average genus for a given
volume, and $\rho$ a constant equal to $10^{-3}\eta^{-3}$ and $4\times
10^{-5}\eta^{-3}$ for vortex clusters and Q-structures respectively.

From relation $\langle g \rangle = \rho V_o$, the genus may be
understood as an alternative method to characterize the level of
complexity of the structures, with $\rho$ a density equal to the
number of holes per unit volume.  If we define $l$ as the average
distance between holes within the structures, its value may be
approximated as $l\approx(10^{-3})^{-1/\alpha_c}\eta \approx 30\eta$
for vortex clusters and $l\approx(4\times 10^{-5})^{-1/\alpha_Q}\eta
\approx 90\eta$ for Q-structures, with $\alpha_c=2$ and
$\alpha_Q=2.25$ the average fractal dimensions of the objects computed
by \citeN{loz:flo:jim:2012}. These lengths are consistent with a model
of coherent structures built by small blocks of length $30-90\eta$
stacked together to create larger objects but not perfectly compacted,
which results in holes between the blocks. For a given volume, $V_o$,
vortex clusters have on average 25 times more holes than Q-structures,
suggesting that their blocks and connections are fundamentally
different.  This is consistent with \citeN{loz:flo:jim:2012} who
showed that the Q-structures are flake-shaped while vortex clusters
are worm-shaped, also visible in Figure
\ref{fig:applications:coherent}.

%--------------------------------------------------------------------------%
\subsection{Turbulent/non-turbulent interface detection in a turbulent jet}
\label{subsec:applications:tnt}
%--------------------------------------------------------------------------%

We use a direct numerical simulation of a time-decaying turbulent jet
(see \citeN[Chapter~5]{pope:2000}) by \citeN{vel:bor:2014} to identify
a turbulent/non-turbulent interface.  A brief introduction about such
interface is presented next.

Two regions can be distinguished in an unbounded turbulent flow, the
fully turbulent region, characterized by strong fluctuations, and the
irrotational free stream. These two regions are, in most cases,
separated by a single thin layer, called turbulent/non-turbulent
interface. The first step to analyze the physical processes that
happen within this interface layer is to locate it. This interface is
known to be fractal-like \cite{Sre:1989} and it contains all the
scales between the smallest and the largest possible. Such a wide
range of scales imposes a strong restriction on the size of the domain
that has to be studied, since small portions would only give reliable
results for the small scales. The most common method to locate the
turbulent/non-turbulent interface is to threshold a scalar field where
the two characteristic states of the flow can be easily
distinguished. \citeN{Sre:1989} and \citeN{FLM:5919136} use the
concentration of a passive scalar injected in the turbulent side, and
threshold it at the least probable value of the concentration.
\citeN{FLM:95049} and \citeN{SilvaTaveira} use a particular isocontour
of the magnitude of vorticity $|\omega|(x,y,z) = \omega_0$, where
vorticity is defined as the rotational of the velocity vector, $\vec
\omega=\nabla \wedge \vec u$. \citeN{gampert2014vorticity} found that
the isocontours obtained thresholding concentration and vorticity
magnitude are similar, and \citeN{dasilva2014characteristics} found
that the least probable value of vorticity magnitude can be used
successfully as a threshold for a variety of turbulent flows. Despite
the convergence of some popular methodologies, other authors like
\citeN{FLM:9176488} have proposed alternative strategies.
%
%_________________________________________________________________%
\begin{figure}
  \centerline{
    \mylab{2cm}{6.4cm}{(\aaa)}
    \includegraphics[width=0.32\textwidth, clip=true, trim=12cm 0cm 12cm 0cm]{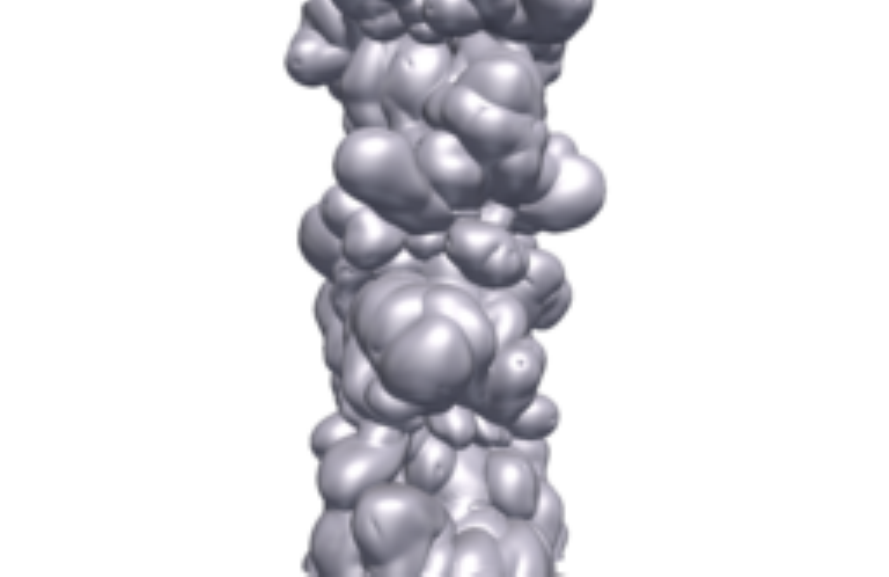}
    \mylab{2cm}{6.4cm}{(\bbb)}
    \includegraphics[width=0.32\textwidth, clip=true, trim=12cm 0cm 12cm 0cm]{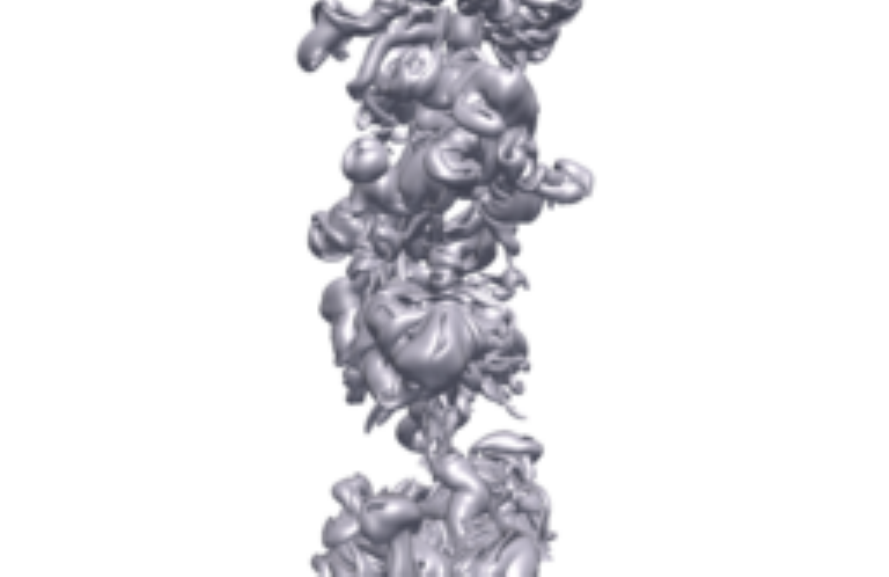}
    \mylab{2cm}{6.4cm}{(\ccc)}
    \includegraphics[width=0.32\textwidth, clip=true, trim=12cm 0cm 12cm 0cm]{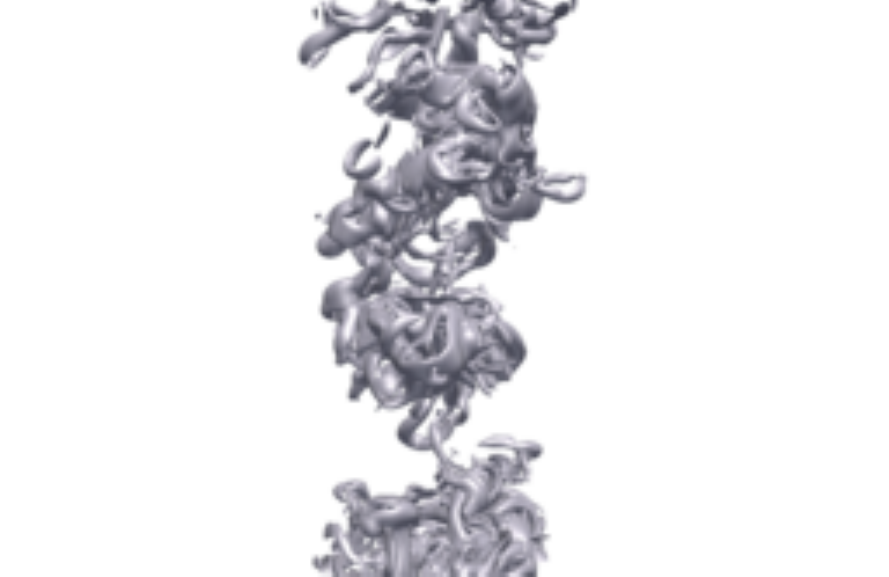}
  }
  \caption{\label{fig:applications:various_tnt} Isocontours of vorticity
    magnitude $|\omega|$, at different thresholds for a turbulent jet.
    Figure (a) corresponds to a very low value of the threshold
    $|\omega|/\omega_{rms}=0.05$, and very low genus, where
    $\omega_{rms}$ is the rooted-mean-squared vorticity
    magnitude. Figure (b) corresponds to the threshold that maximize the
    genus $|\omega|/\omega_{rms}=3$, and (c) to a threshold slightly
    higher than (b) $|\omega|/\omega_{rms}=5$.}
\end{figure}
%_________________________________________________________________%

One important aspect of the choice of the threshold is the impact it
has on the geometry of the interface. If the threshold $\omega_0$ is a
low value of vorticity, like the detection shown in Figure
\ref{fig:applications:various_tnt}(a), the interface is relatively
simple, showing that the perturbation caused by the turbulent motion
is smoothed out farther down the free stream. On the other hand, as
soon as the threshold is slightly increased, the surface is populated
with a large amount of handles (or holes), as can be seen in Figures
\ref{fig:applications:various_tnt}(b,c).  These handles are most
likely a geometrical feature of the fully turbulent flow. Depending on
the value of the threshold, the surface generated has different
topological properties.

\begin{figure}
  \centerline{
    \includegraphics[width=0.7\textwidth]{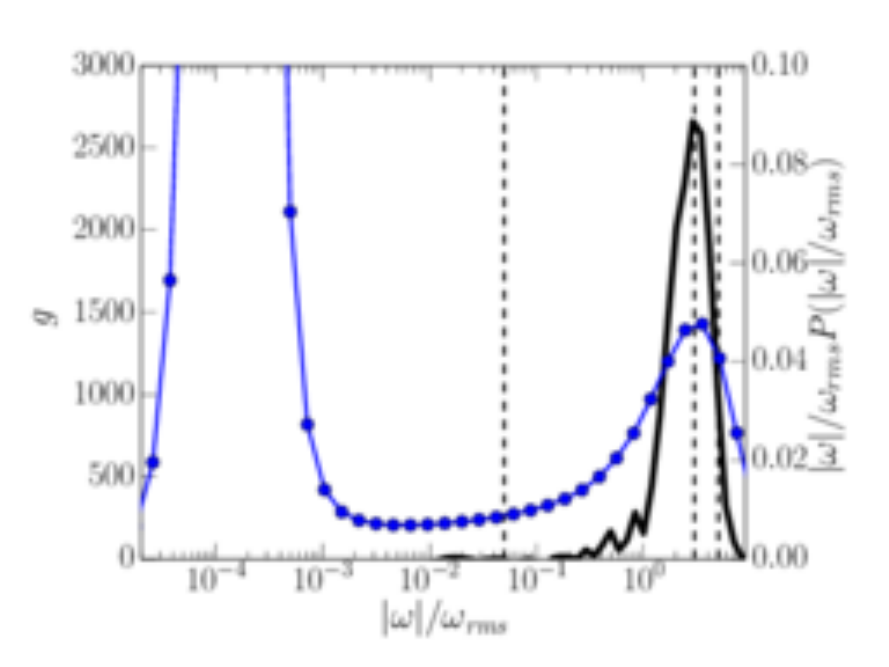}}
  \caption{\label{fig:applications:genus_tnt} (Blue line with circles)
    Premultiplied PDF of vorticity magnitude $|\omega|$ normalized with its
    root-mean-squared value in a turbulent temporal round jet. (Solid black
    line) Genus of the turbulent/non-turbulent interface as a function of the
    vorticity magnitude. The vertical dashed lines are the thresholds used in
    Figures \ref{fig:applications:various_tnt}(a,b,c).}
\end{figure}

The geometrical complexity, measured in this case with the number
handles, has an important side effect on the analysis of the
properties of the flow depending on the relative position to the
interface. Two relatively popular assumptions about the interface are
that there is a privileged direction across which the relative
distance to the interface can be measured
\cite{FLM:5919136,SilvaTaveira}, and that the interface is simple
enough so that a local normal is meaningful
\cite{FLM:95049,FLM:9176488}. These two assumptions are not strictly
correct if handles are a dominant feature of the interface. At the
same time, the criterion explored by
\citeN{dasilva2014characteristics} depends on the characteristics of
the non-turbulent region. The probability density function of
vorticity in the same round jet of Figure
\ref{fig:applications:various_tnt} is shown in Figure
\ref{fig:applications:genus_tnt}. It has been premultiplied to
emphasize the fact that the PDF has two major contributions, one
from the bulk of the non-turbulent flow with low vorticity (left
peak), and a second one from the bulk of the turbulent flow with high
vorticity (right peak). Note that, if the flow was in an ideal state
with no perturbations, the left peak would be in in the limit of
vanishing vorticity. In consequence, the outcome of the criterion
defined by \cite{Sre:1989} applied to the vorticity field can be
intuitively defined as the lowest threshold that is not affected by
the spurious vorticity present in the free stream. This criterion is
strictly correct, but it may be more representative of the smoothed
out perturbations relatively far from the turbulent motion.  It is
therefore necessary to explore other complementary threshold choices
that provide a more complete description of the vorticity field.
%
%_________________________________________________________________%
%_________________________________________________________________%

The genus of the surface detected as a function of the value used to
threshold the vorticity magnitude is presented in Figure
\ref{fig:applications:genus_tnt}.  The results have been averaged
using an ensemble of four equivalent cases. The curve shows that there
is a gradual yet evident change in the topological properties of the
vorticity interface from a threshold $\omega_0\sim\omega_{rms}$.
Beyond that value, handles are a dominant feature of the interface,
and the standard tools for the conditional analysis are probably not
valid. If the criterion of minimum probability provides a lower limit
for the threshold, the genus of the interface is an useful criterion
for an upper limit.

%=================================================================%
\section{Conclusions}\label{sec:conclusions}
%=================================================================%

We have presented and validated a simple algorithm to numerically
compute the genus of discrete surfaces using the Euler characteristic
formula. The method is valid for surfaces associated with
three-dimensional objects obtained by thresholding a discrete scalar
field defined in a structured-collocated grid and offers several
advantages. First, it does not rely on any direct triangulation of the
surfaces, which is usually memory and time-consuming.  Besides, the
surfaces of all the 3D objects in the domain are automatically
detected and the genus is exactly computed without any spurious
holes. Last, but not least, it needs practically zero memory, it is
fast and scalable, with a computational cost directly proportional to
the size of the grid computed. The algorithm is also highly
parallelizable, and a Fortran Coarrays version was implemented to take
advantage of multicore processors without increasing the memory usage.
This makes the algorithm suitable for large datasets, like the ones
encountered in direct numerical simulations of turbulent flows.  Two
applications to the characterization of complex structures in
turbulent flows have been presented. In the first case, the genus of
coherent structures extracted from a turbulent channel flows is
computed and found to be proportional to the volume of the objects.
In the second application, the genus is used to find an appropriate
threshold to detect the turbulent/non-turbulent interface in a
turbulent jet.

% Acknowledgments
\begin{acks}
The authors would like to acknowledge fruitful discussions with Javier
Jim\'enez and Jos\'e Cardesa-Due\~nas, and to thank Profs. Dolors
Ayala and Irving Cruz for providing the test data used in the early
stages of the work.  We are also very grateful to Tim Hopkins for his
assistance with the software component of the paper, and to the
referees for their very constructive feedback.
\end{acks}

% Bibliography
\bibliographystyle{ACM-Reference-Format-Journals}
\bibliography{genus}

\end{document}